\documentclass[10pt]{article}
\usepackage{graphicx}
\usepackage{amsmath}
\usepackage{amssymb}
\usepackage{caption2}
\begin{document}
\begin{center}
{\large {\bf \sc{  Strong decays of the $Y(4660)$ as a vector tetraquark state in solid quark-hadron duality
  }}} \\[2mm]
Zhi-Gang  Wang \footnote{E-mail: zgwang@aliyun.com.  }    \\
 Department of Physics, North China Electric Power University, Baoding 071003, P. R. China
\end{center}

\begin{abstract}
In this article, we  choose  the $[sc]_P[\bar{s}\bar{c}]_A-[sc]_A[\bar{s}\bar{c}]_P$  type tetraquark current to study the hadronic coupling constants in the strong decays $Y(4660)\to J/\psi f_0(980)$, $ \eta_c \phi(1020)$,    $ \chi_{c0}\phi(1020)$, $ D_s \bar{D}_s$, $ D_s^* \bar{D}^*_s$, $ D_s \bar{D}^*_s$,  $ D_s^* \bar{D}_s$, $  \psi^\prime \pi^+\pi^-$, $J/\psi\phi(1020)$ with the QCD sum rules based on solid quark-hadron quality. The predicted width $\Gamma(Y(4660) )= 74.2^{+29.2}_{-19.2}\,{\rm{MeV}}$ is in excellent agreement with the experimental data $68\pm 11\pm 1 {\mbox{ MeV}}$ from the Belle collaboration, which supports assigning the $Y(4660)$ to be the  $[sc]_P[\bar{s}\bar{c}]_A-[sc]_A[\bar{s}\bar{c}]_P$  type tetraquark state with $J^{PC}=1^{--}$.
In calculations, we observe that the  hadronic coupling constants $ |G_{Y\psi^\prime f_0}|\gg |G_{Y J/\psi f_0}|$, which is consistent with the observation of the $Y(4660)$ in the $\psi^\prime\pi^+\pi^-$ mass spectrum, and favors the $\psi^{\prime}f_0(980)$ molecule assignment.  It is important to search for the process $Y(4660)\to J/\psi \phi(1020)$ to diagnose the nature  of the $Y(4660)$, as the decay is greatly suppressed.

\end{abstract}



\section{Introduction}

In 2007,  the Belle collaboration  observed the $Y(4360)$ and $Y(4660)$ in the $\pi^+ \pi^- \psi^\prime$ invariant mass distribution with statistical significances $8.0\sigma$ and $5.8\sigma$ respectively in the precess  $e^+e^- \to \gamma_{\rm ISR}\pi^+ \pi^- \psi^\prime$ between threshold and $\sqrt{s}=5.5 \,\rm{GeV}$  using $673 \rm{fb}^{-1}$ of data collected with the Belle detector at KEKB \cite{Belle-Y4660-2007}.
In 2008, the Belle collaboration  observed the $Y(4630)$ in the $\Lambda_c^+ \Lambda_c^-$ invariant mass distribution with a significance of $8.2\sigma$
 in the exclusive process $e^+e^- \to \gamma_{\rm ISR} \Lambda_c^+ \Lambda_c^-$  with an integrated luminosity of $695 \rm{fb}^{-1}$ at the KEKB \cite{Belle-Y4630-2008}.
  The values of the mass and width of the $Y(4630)$ are  consistent within errors with that of
a new charmonium-like state $Y(4660)$.

In 2014, the Belle collaboration measured  the $e^+e^- \to \gamma_{\rm ISR}\pi^+ \pi^- \psi^\prime$ cross section  from $4.0$ to $5.5\,\rm{ GeV}$ with
the full data sample of the Belle experiment using the ISR (initial state radiation) technique, and determined the parameters of the
$Y(4360)$ and $Y(4660)$ resonances and superseded previous Belle determination \cite{Belle-Y4660-2014}. The masses and widths are shown explicitly in Table 1.
Furthermore, the Belle collaboration studied the
$\pi^+\pi^-$ invariant mass distribution and observed that there are two clusters of events around the masses of the $f_0(500)$ and $f_0(980)$ corresponding to the $Y(4360)$ and $Y(4660)$,  respectively.
The $J^{PC}$ quantum numbers of the final states accompanying
the ISR photon(s) are restricted to $J^{PC}=1^{--}$.  According to potential
model calculations \cite{cc-spectrum,KTChao-4660}, the $4^3{\rm S}_1$, $5^3{\rm S}_1$, $6^3{\rm S}_1$ and $3^3{\rm D}_1$ charmonium states are expected to
be in the mass range close to the two resonances $Y(4360)$ and $Y(4660)$, however,  there are no enough
vector charmonium candidates  which can match those new $Y$ states consistently.

Now, let us begin with discussing the nature of the $f_0(500)$ and $f_0(980)$ to explore the $Y(4660)$.  In the  scenario of conventional two-quark  states, the structures of the $f_0(500)$ and $f_0(980)$
 in the ideal mixing limit can be symbolically written as,
\begin{eqnarray}
f_0(500)= \frac{\bar{u}u+\bar{d}d}{\sqrt{2}}\, ,\;f_0(980)= \bar{s}s\,  .
\end{eqnarray}
While in the  scenario of tetraquark  states, the structures of the $f_0(500)$ and $f_0(980)$ in the ideal mixing limit can be symbolically written as
\cite{Close2002,ReviewAmsler2,Maiani-Scalar},
\begin{eqnarray}
f_0(500)=ud\bar{u}\bar{d}\, ,\; f_0(980)={us\bar{u}\bar{s}+ds\bar{d}\bar{s}\over\sqrt{2}}\, .
\end{eqnarray}
In Ref.\cite{WangScalarNonet}, we take the nonet scalar mesons  below $1\,\rm{ GeV}$ as the two-quark-tetraquark  mixed states and study their  masses and pole residues  with the  QCD sum rules in details.   We  determine the mixing angles, which indicate that the dominant components are the two-quark components. The $Y(4660)$ maybe have  $\bar{s}s$ constituent. The decay $Y(4630)\to \Lambda_c^+ \Lambda_c^-$ has been observed, if the $Y(4660)$ and $Y(4630)$ are the same particle, the decay $Y(4630)\to \Lambda_c^+ \Lambda_c^-$ is   Okubo-Zweig-Iizuka suppressed,  there should be some rescattering mechanism  to account for the decay.

The threshold of the $\psi^\prime f_0(980)$ is $4676\,\rm{MeV}$ from the Particle Data Group \cite{PDG}, which is just above the mass $m_{Y(4660)}=4652\pm10\pm 8\,\rm{MeV}$ from the Belle collaboration \cite{Belle-Y4660-2014}. The $Y(4660)$ can be  assigned to be a $\psi^\prime f_0(980)$ molecular state \cite{FKGuo-4660,Wang-CTP-4660,Nielsen-4660-mole} or a $\psi^\prime f_0(980)$ hadro-charmonium \cite{Hadro-Charm}. Other assignments, such as  a 2P $[cq]_S[\bar{c}\bar{q}]_S$ tetraquark state \cite{Ebert-4660-2P}, a $\psi({\rm 6S })$ state \cite{KTChao-4660},  a $\psi({\rm 5S })$ state \cite{XHZhong-Y4660},  a ground state P-wave tetraquark state \cite{Nielsen-4260-4660,Azizi-4660,ChenZhu,Wang-tetra-formula,WangY4360Y4660-1803,WangEPJC-1601,ZhangHuang-PRD} are also possible.

In Table 2, we list out the predictions of the masses of the vector tetraquark (tetraquark molecule) states based on the QCD sum rules \cite{Wang-CTP-4660,Nielsen-4660-mole,Nielsen-4260-4660,Azizi-4660,ChenZhu,Wang-tetra-formula,WangY4360Y4660-1803,WangEPJC-1601,ZhangHuang-PRD}, where the $S$, $P$, $A$ and $V$ denote the scalar ($S$), pseudoscalar ($P$), axialvector ($A$) and vector ($V$) diquark states. From the Table, we can see that it is not difficult to reproduce the experimental value of the mass of the $Y(4660)$ with the QCD sum rules. However, the  quantitative   predications depend on the quark structures,
the input parameters at the QCD side, the pole contributions of the ground states, and the truncations of  the operator product expansion.

In the QCD sum rules for the hidden-charm (or hidden-bottom) tetraquark states and molecular states, the integrals
 \begin{eqnarray}
 \int_{4m_Q^2(\mu)}^{s_0} ds \rho_{QCD}(s,\mu)\exp\left(-\frac{s}{T^2} \right)\, ,
 \end{eqnarray}
are sensitive to the energy scales $\mu$, where the $\rho_{QCD}(s,\mu)$ are  the QCD spectral densities, the $T^2$ are the Borel parameters, the $s_0$ are the continuum thresholds parameters, the predicted masses depend heavily on the energy scales $\mu$.
In Refs.\cite{Wang-tetra-formula,Wang-tetra-NPA,WangHuang-molecule}, we suggest an energy scale formula $\mu=\sqrt{M^2_{X/Y/Z}-(2{\mathbb{M}}_Q)^2}$ with the effective $Q$-quark mass ${\mathbb{M}}_Q$ to determine the ideal energy scales of the QCD spectral densities. The formula enhances the pole contributions remarkably, we obtain the pole contributions as large as $(40-60)\%$, the largest pole contributions up to now. Compared to the old values obtained  in Ref.\cite{Wang-tetra-formula}, the new values based on detailed analysis  with the   updated   parameters  are preferred \cite{WangY4360Y4660-1803}.
The energy scale formula also works well in the QCD sum rules  for the hidden-charm pentaquark states \cite{Wang1508-EPJC}.

For the  correlation functions  of the hidden-charm (or hidden-bottom) tetraquark currents,  there are two heavy quark propagators and two light quark propagators, if each heavy quark line emits a gluon and each light quark line contributes  a quark pair, we obtain a operator $GG\bar{q}q\bar{q}q$, which is of dimension $10$,   we should take into account the vacuum condensates at least up to dimension $10$ in the operator product expansion.

In Refs.\cite{Wang-tetra-formula,WangY4360Y4660-1803,WangEPJC-1601,WZG1809}, we study the mass spectrum of the vector tetraquark states in a comprehensive  way by
carrying out the operator product expansion up to the vacuum condensates of dimension  $10$, and use the  energy scale formula $\mu=\sqrt{M^2_{X/Y/Z}-(2{\mathbb{M}}_c)^2}$ or modified energy scale formula $\mu=\sqrt{M^2_{X/Y/Z}-(2{\mathbb{M}}_c+0.5\,\rm{GeV})^2}=\sqrt{M^2_{X/Y/Z}-(4.1\,\rm{GeV})^2}$  to determine the ideal  energy scales of the QCD spectral densities in a consistent  way. In the  scenario of tetraquark  states, we observe that the preferred quark configurations for the $Y(4660)$ are the $[sc]_P[\bar{s}\bar{c}]_A-[sc]_A[\bar{s}\bar{c}]_P$  and $[qc]_A[\bar{q}\bar{c}]_A$. In this article, we choose the quark configuration $[sc]_P[\bar{s}\bar{c}]_A-[sc]_A[\bar{s}\bar{c}]_P$ to  examine the nature of the $Y(4660)$.

In Ref.\cite{WangZhang-Solid}, we assign  the $Z_c^\pm(3900)$ to be  the diquark-antidiquark  type axialvector  tetraquark  state,
 study the hadronic coupling  constants $G_{Z_cJ/\psi\pi}$, $G_{Z_c\eta_c\rho}$, $G_{Z_cD \bar{D}^{*}}$ with the QCD sum rules  by taking
  into account both the connected and disconnected Feynman diagrams in the operator product expansion. We pay special attentions to matching the hadron side of the correlation functions with the QCD side of the correlation functions to obtain solid duality. The routine works well in studying the decays $X(4140/4274) \to J/\psi\phi(1020)$ \cite{Wang-Y4140-Y4274}.

In this article, we assign  the $Y(4660)$ to be  the $[sc]_P[\bar{s}\bar{c}]_A-[sc]_A[\bar{s}\bar{c}]_P$ type vector tetraquark state, and study the strong decays
$Y(4660)\to J/\psi f_0(980)$, $ \eta_c \phi(1020)$,    $ \chi_{c0}\phi(1020)$, $ D_s \bar{D}_s$, $ D_s^* \bar{D}^*_s$,
$ D_s \bar{D}^*_s$, $ D_s^* \bar{D}_s$,  $\psi^\prime \pi^+\pi^-$, $J/\psi\phi(1020)$ with the QCD sum rules based on the solid quark-hadron duality, and reexamine the assignment of the $Y(4660)$.

 \begin{table}
\begin{center}
\begin{tabular}{|c|c|c|c|c|c|c|c|}\hline\hline

Year  &           &Mass (MeV)                    &Width (MeV)                           &Experiment     \\ \hline\hline

2007  &$Y(4360)$  &$4361\pm 9\pm 9$              &$74\pm 15\pm 10 $                     &Belle \cite{Belle-Y4660-2007}     \\
      &$Y(4660)$  &$4664\pm 11\pm 5$             &$48\pm 15\pm 3$                       &Belle \cite{Belle-Y4660-2007}    \\  \hline

2008  &$Y(4630)$  &$4634^{+8}_{-7}{}^{+5}_{-8}$  &$92^{+40}_{-24}{}^{+10}_{-21}$        &Belle \cite{Belle-Y4630-2008}    \\  \hline

2014  &$Y(4360)$  &$4347\pm 6\pm 3$              &$103\pm 9\pm 5 $                      &Belle \cite{Belle-Y4660-2014}     \\
      &$Y(4660)$  &$4652\pm10\pm 8$              &$68\pm 11\pm 1$                       &Belle \cite{Belle-Y4660-2014}    \\  \hline\hline
 \end{tabular}
\end{center}
\caption{ The  masses and widths from the different experiments. }
\end{table}

\begin{table}
\begin{center}
\begin{tabular}{|c|c|c|c|c|c|c|c|}\hline\hline
           &Structures                                                      &OPE\,(No)    &mass(GeV)      &References   \\ \hline

$Y(4660)$  &$\psi^\prime f_0(980)$                                          &$10$         &$4.71$         &\cite{Wang-CTP-4660}  \\
$Y(4660)$  &$\psi^\prime f_0(980)$                                          &$6$          &$4.67$         &\cite{Nielsen-4660-mole}  \\ \hline

$Y(4660)$  &$[sc]_S[\bar{s}\bar{c}]_V+[sc]_V[\bar{s}\bar{c}]_S$             &$8\,(7)$     &$4.65$         &\cite{Nielsen-4260-4660}  \\
$Y(4660)$  &$[sc]_S[\bar{s}\bar{c}]_V+[sc]_V[\bar{s}\bar{c}]_S$             &$10$         &$4.68$         &\cite{Azizi-4660}  \\  \hline

$Y(4660)$  &$[qc]_S[\bar{q}\bar{c}]_V+[qc]_V[\bar{q}\bar{c}]_S$             &$8\,(7)$     &$4.64$         &\cite{ChenZhu}  \\
$Y(4360)$  &$[qc]_S[\bar{q}\bar{c}]_V+[qc]_V[\bar{q}\bar{c}]_S$             &$10$         &$4.34$         &\cite{WangY4360Y4660-1803}  \\ \hline

$Y(4660)$  &$[sc]_P[\bar{s}\bar{c}]_A-[sc]_A[\bar{s}\bar{c}]_P$             &$10$         &$4.70$         &\cite{Wang-tetra-formula}  \\
$Y(4660)$  &$[sc]_P[\bar{s}\bar{c}]_A-[sc]_A[\bar{s}\bar{c}]_P$             &$10$         &$4.66$         &\cite{WangY4360Y4660-1803}  \\ \hline

$Y(4660)$  &$[qc]_P[\bar{q}\bar{c}]_A-[qc]_A[\bar{q}\bar{c}]_P$             &$10$         &$4.66$         &\cite{Wang-tetra-formula}   \\
$Y(4660)$  &$[qc]_P[\bar{q}\bar{c}]_A-[qc]_A[\bar{q}\bar{c}]_P$             &$10$         &$4.59$         &\cite{WangY4360Y4660-1803}  \\  \hline

$Y(4660)$  &$[qc]_A[\bar{q}\bar{c}]_A$                                      &$10$         &$4.66$         &\cite{WangEPJC-1601}  \\ \hline

$Y(4660)$  &$[sc]_S[\bar{s}\bar{c}]_S$                                      &$6$          &$4.69$         &\cite{ZhangHuang-PRD}  \\ \hline   \hline
\end{tabular}
\end{center}
\caption{ The  masses from the QCD sum rules with different quark structures, where the OPE denotes  truncations of  the operator product expansion up to the vacuum condensates of dimension $n$, the No denotes the vacuum condensates of dimension $n^\prime$ are not included.   }
\end{table}

The article is arranged as follows:  we illustrate how to calculate the hadronic coupling constants in the two-body strong
 decays of the tetraquark states with the QCD sum rules   in section 2,  in section 3, we obtain the QCD sum rules for  the hadronic coupling constants
 $G_{Y J/\psi f_0}$, $ G_{Y\eta_c \phi}$,    $ G_{Y\chi_{c0}\phi}$, $ G_{YD_s \bar{D}_s}$, $G_{Y D_s^* \bar{D}^*_s}$,
$ G_{YD_s \bar{D}^*_s}$,   $ G_{Y\psi^\prime f_0}$, $G_{YJ/\psi\phi}$;  section 4 is reserved for our conclusion.

\section{The hadronic coupling constants in the two-body strong  decays of the tetraquark states  }

In this section, we illustrate how to calculate the hadronic coupling constants in the two-body strong
 decays of the tetraquark states with the QCD sum rules.
We write down the  three-point correlation functions $\Pi(p,q)$ firstly,
\begin{eqnarray}
\Pi(p,q)&=&i^2\int d^4xd^4y e^{ipx}e^{iqy}\langle 0|T\left\{J_{B}(x)J_{C}(y)J_{A}^{\dagger}(0)\right\}|0\rangle\, ,
\end{eqnarray}
where the currents $J_A(0)$ interpolate the tetraquark states $A$, the $J_B(x)$ and $J_C(y)$ interpolate the  conventional mesons $B$ and $C$,  respectively,
\begin{eqnarray}
\langle0|J_{A}(0)|A(p^\prime)\rangle&=&\lambda_{A} \,\, , \nonumber \\
\langle0|J_{B}(0)|B(p)\rangle&=&\lambda_{B} \,\, , \nonumber \\
\langle0|J_{C}(0)|C(q)\rangle&=&\lambda_{C} \,\, ,
\end{eqnarray}
the $\lambda_A$, $\lambda_B$  and $\lambda_{C}$ are the pole residues or   decay constants.

At the phenomenological side,  we insert  a complete set of intermediate hadronic states with
the same quantum numbers as the current operators $J_A(0)$, $J_B(x)$, $J_{C}(y)$ into the three-point
correlation functions $\Pi(p,q)$ and  isolate the ground state
contributions to obtain the  result \cite{SVZ79,Reinders85},
\begin{eqnarray}
\Pi(p,q)&=& \frac{\lambda_{A}\lambda_{B}\lambda_{C}G_{ABC} }{(m_{A}^2-p^{\prime2})(m_{B}^2-p^2)(m_{C}^2-q^2)}+ \frac{1}{(m_{A}^2-p^{\prime2})(m_{B}^2-p^2)} \int_{s^0_C}^\infty dt\frac{\rho_{AC^\prime}(p^{\prime 2},p^2,t)}{t-q^2}\nonumber\\
&& + \frac{1}{(m_{A}^2-p^{\prime2})(m_{C}^2-q^2)} \int_{s^0_{B}}^\infty dt\frac{\rho_{AB^\prime}(p^{\prime 2},t,q^2)}{t-p^2}  \nonumber\\
&& + \frac{1}{(m_{B}^2-p^{2})(m_{C}^2-q^2)} \int_{s^0_{A}}^\infty dt\frac{\rho_{A^{\prime}B}(t,p^2,q^2)+\rho_{A^{\prime}C}(t,p^2,q^2)}{t-p^{\prime2}}+\cdots \nonumber\\
&=&\Pi(p^{\prime2},p^2,q^2) \, ,
\end{eqnarray}
where $p^\prime=p+q$,  the $G_{ABC}$  are the hadronic coupling constants defined by
\begin{eqnarray}
\langle B(p)C(q)|A(p^{\prime})\rangle&=&i G_{ABC}  \, ,
\end{eqnarray}
the four   functions $\rho_{AC^\prime}(p^{\prime 2},p^2,t)$, $ \rho_{AB^\prime}(p^{\prime 2},t,q^2)$,
$ \rho_{A^{\prime}B}(t^\prime,p^2,q^2)$ and $\rho_{A^{\prime}C}(t^\prime,p^2,q^2)$
   have complex dependence on the transitions between the ground states and the higher resonances  or the continuum states.

We rewrite the correlation functions  $\Pi_H(p^{\prime 2},p^2,q^2)$ at the hadron  side as
\begin{eqnarray}
\Pi_{H}(p^{\prime 2},p^2,q^2)&=&\int_{(m_{B}+m_{C})^2}^{s_{A}^0}ds^\prime \int_{\Delta_s^2}^{s^0_{B}}ds \int_{\Delta_u^2}^{u^0_{C}}du  \frac{\rho_H(s^\prime,s,u)}{(s^\prime-p^{\prime2})(s-p^2)(u-q^2)}\nonumber\\
&&+\int_{s^0_A}^{\infty}ds^\prime \int_{\Delta_s^2}^{s^0_{B}}ds \int_{\Delta_u^2}^{u^0_{C}}du  \frac{\rho_H(s^\prime,s,u)}{(s^\prime-p^{\prime2})(s-p^2)(u-q^2)}+\cdots\, ,
\end{eqnarray}
 through dispersion relation, where the $\rho_H(s^\prime,s,u)$   are the hadronic spectral densities,
\begin{eqnarray}
\rho_H(s^\prime,s,u)&=&{\lim_{\epsilon_3\to 0}}\,\,{\lim_{\epsilon_2\to 0}} \,\,{\lim_{\epsilon_1\to 0}}\,\,\frac{ {\rm Im}_{s^\prime}\, {\rm Im}_{s}\,{\rm Im}_{u}\,\Pi_H(s^\prime+i\epsilon_3,s+i\epsilon_2,u+i\epsilon_1) }{\pi^3} \, ,
\end{eqnarray}
where the $\Delta_s^2$ and $\Delta_u^2$ are the thresholds, the  $s_{A}^0$, $s_{B}^0$, $u_{C}^0$ are the continuum thresholds.

Now we carry out the operator product expansion at the QCD side, and write the correlation functions  $\Pi_{QCD}(p^{\prime 2},p^2,q^2)$  as
\begin{eqnarray}
\Pi_{QCD}(p^{\prime 2},p^2,q^2)&=&  \int_{\Delta_s^2}^{s^0_{B}}ds \int_{\Delta_u^2}^{u^0_{C}}du  \frac{\rho_{QCD}(p^{\prime2},s,u)}{(s-p^2)(u-q^2)}+\cdots\, ,
\end{eqnarray}
through dispersion relation, where the $\rho_{QCD}(p^{\prime 2},s,u)$   are the QCD spectral densities,
\begin{eqnarray}
\rho_{QCD}(p^{\prime 2},s,u)&=& {\lim_{\epsilon_2\to 0}} \,\,{\lim_{\epsilon_1\to 0}}\,\,\frac{  {\rm Im}_{s}\,{\rm Im}_{u}\,\Pi_{QCD}(p^{\prime 2},s+i\epsilon_2,u+i\epsilon_1) }{\pi^2} \, .
\end{eqnarray}

However,   the QCD spectral densities $\rho_{QCD}(s^\prime,s,u)$ do  not exist,
\begin{eqnarray}
\rho_{QCD}(s^\prime,s,u)&=&{\lim_{\epsilon_3\to 0}}\,\,{\lim_{\epsilon_2\to 0}} \,\,{\lim_{\epsilon_1\to 0}}\,\,\frac{ {\rm Im}_{s^\prime}\, {\rm Im}_{s}\,{\rm Im}_{u}\,\Pi_{QCD}(s^\prime+i\epsilon_3,s+i\epsilon_2,u+i\epsilon_1) }{\pi^3} \nonumber\\
&=&0\, ,
\end{eqnarray}
because
\begin{eqnarray}
{\lim_{\epsilon_3\to 0}}\,\,\frac{ {\rm Im}_{s^\prime}\,\Pi_{QCD}(s^\prime+i\epsilon_3,p^2,q^2) }{\pi} &=&0\, .
\end{eqnarray}
Thereafter we will write the QCD spectral densities  $\rho_{QCD}(p^{\prime 2},s,u)$ as $\rho_{QCD}(s,u)$ for simplicity.

We math the hadron side of the correlation functions  with the QCD side of the correlation functions,
and carry out the integral over $ds^\prime$  firstly to obtain the solid duality \cite{WangZhang-Solid},
\begin{eqnarray}
\int_{\Delta_s^2}^{s_B^0} ds \int_{\Delta_u^2}^{u_C^0} du \frac{\rho_{QCD}(s,u)}{(s-p^2)(u-q^2)}&=&\int_{\Delta_s^2}^{s_B^0} ds \int_{\Delta_u^2}^{u_C^0} du \frac{1}{(s-p^2)(u-q^2)}\left[ \int_{\Delta^2}^{\infty} ds^\prime \frac{\rho_{H}(s^{\prime},s,u)}{s^\prime-p^{\prime2}}\right]\, , \nonumber\\
\end{eqnarray}
 the  $\Delta^2$ denotes the thresholds $(m_{B}+m_{C})^2$.
 Now we write down  the quark-hadron duality explicitly,
 \begin{eqnarray}
  \int_{\Delta_c^2}^{s^0_{B}}ds \int_{\Delta_u^2}^{u^0_{C}}du  \frac{\rho_{QCD}(s,u)}{(s-p^2)(u-q^2)}&=& \int_{\Delta_c^2}^{s^0_{B}}ds \int_{\Delta_u^2}^{u^0_{C}}du   \int_{(m_{B}+m_{C})^2}^{\infty}ds^\prime \frac{\rho_H(s^\prime,s,u)}{(s^\prime-p^{\prime2})(s-p^2)(u-q^2)} \nonumber\\
  &=&\frac{\lambda_{A}\lambda_{B}\lambda_{C}G_{ABC} }{(m_{A}^2-p^{\prime2})(m_{B}^2-p^2)(m_{C}^2-q^2)} +\frac{C_{A^{\prime}B}+C_{A^{\prime}C}}{(m_{B}^2-p^{2})(m_{C}^2-q^2)} \, . \nonumber\\
\end{eqnarray}
 No approximation is needed, we do not need the continuum threshold parameter $s^0_{A}$ in the $s^\prime$ channel. The $s^\prime$ channel and $s$ channel
  are quite different, we can not set the continuum threshold parameters in the $s$ channel as $s_B^0=s_A^0$, i.e. we can not set $s_B^0=s_{Y}^0=\left(5.15\,\rm{GeV}\right)^2$ in the present case, where the $B$ denotes  the $J/\psi$, $\eta_c$, $\bar{D}_s$, $\bar{D}_s^*$, because the
  contaminations from the excited states $\psi^\prime$, $\eta_c^\prime$, $\bar{D}_s^\prime$, $\bar{D}_s^{*\prime}$ are out of control.

  We can introduce the parameters $C_{AC^\prime}$, $C_{AB^\prime}$, $C_{A^\prime B}$ and $C_{A^\prime C}$   to parameterize the net effects,
\begin{eqnarray}
C_{AC^\prime}&=&\int_{s^0_C}^\infty dt\frac{ \rho_{AC\prime}(p^{\prime 2},p^2,t)}{t-q^2}\, ,\nonumber\\
C_{AB^\prime}&=&\int_{s^0_{B}}^\infty dt\frac{\rho_{AB^\prime}(p^{\prime 2},t,q^2)}{t-p^2}\, ,\nonumber\\
C_{A^\prime B}&=&\int_{s^0_{A}}^\infty dt\frac{ \rho_{A^\prime B}(t,p^2,q^2)}{t-p^{\prime2}}\, ,\nonumber\\
C_{A^\prime C}&=&\int_{s^0_{A}}^\infty dt\frac{ \rho_{A^\prime C}(t,p^2,q^2)}{t-p^{\prime2}}\, .
\end{eqnarray}
In numerical calculations,   we   take the relevant functions $C_{A^\prime B}$ and $C_{A^\prime C}$  as free parameters, and choose suitable values  to eliminate the contaminations from the higher resonances and continuum states to obtain the stable QCD sum rules with the variations of
the Borel parameters.

If the  $B$  are charmonium or bottomnium states, we set  $p^{\prime2}=p^2$  and perform the double Borel transform  with respect to the variables $P^2=-p^2$ and $Q^2=-q^2$, respectively  to obtain the  QCD sum rules,
\begin{eqnarray}
&& \frac{\lambda_{A}\lambda_{B}\lambda_{C}G_{ABC}}{m_{A}^2-m_{B}^2} \left[ \exp\left(-\frac{m_{B}^2}{T_1^2} \right)-\exp\left(-\frac{m_{A}^2}{T_1^2} \right)\right]\exp\left(-\frac{m_{C}^2}{T_2^2} \right) +\nonumber\\
&&\left(C_{A^{\prime}B}+C_{A^{\prime}C}\right) \exp\left(-\frac{m_{B}^2}{T_1^2} -\frac{m_{C}^2}{T_2^2} \right)=\int_{\Delta_s^2}^{s_B^0} ds \int_{\Delta_u^2}^{u_C^0} du\, \rho_{QCD}(s,u)\exp\left(-\frac{s}{T_1^2} -\frac{u}{T_2^2} \right)\, ,
\end{eqnarray}
where the $T_1^2$ and $T_2^2$ are the Borel parameters.
If the  $B$   are open-charm or open-bottom mesons, we set  $p^{\prime2}=4p^2$  and perform the double Borel transform  with respect to the variables $P^2=-p^2$ and $Q^2=-q^2$, respectively  to obtain the  QCD sum rules,
\begin{eqnarray}
&& \frac{\lambda_{A}\lambda_{B}\lambda_{C}G_{ABC}}{4\left(\widetilde{m}_{A}^2-m_{B}^2\right)} \left[ \exp\left(-\frac{m_{B}^2}{T_1^2} \right)-\exp\left(-\frac{\widetilde{m}_{A}^2}{T_1^2} \right)\right]\exp\left(-\frac{m_{C}^2}{T_2^2} \right) +\nonumber\\
&&\left(C_{A^{\prime}B}+C_{A^{\prime}C}\right) \exp\left(-\frac{m_{B}^2}{T_1^2} -\frac{m_{C}^2}{T_2^2} \right)=\int_{\Delta_s^2}^{s_B^0} ds \int_{\Delta_u^2}^{u_C^0} du\, \rho_{QCD}(s,u)\exp\left(-\frac{s}{T_1^2} -\frac{u}{T_2^2} \right)\, ,
\end{eqnarray}
where $\widetilde{m}_A^2=\frac{m_A^2}{4}$.

\section{The width of the $Y(4660)$ as a vector tetraquark state }

Now we write down the  three-point correlation functions for the strong decays $Y(4660)\to J/\psi f_0(980)$, $ \eta_c \phi(1020)$,    $ \chi_{c0}\phi(1020)$, $ D_s \bar{D}_s$, $ D_s^* \bar{D}^*_s$, $ D_s \bar{D}^*_s$,  $ D_s^* \bar{D}_s$, $  \psi^\prime \pi^+\pi^-$, $J/\psi\phi(1020)$, respectively, and apply the method presented in previous section to obtain the QCD sum rules for the hadronic coupling constants $G_{Y J/\psi f_0}$,
$ G_{Y\eta_c \phi}$,    $G_{Y\chi_{c0}\phi}$, $G_{YD_s \bar{D}_s}$, $G_{Y D_s^* \bar{D}^*_s}$, $G_{YD_s \bar{D}^*_s}$,   $G_{Y\psi^\prime f_0}$, $G_{YJ/\psi\phi}$.

For the two-body strong decays  $Y(4660)\to J/\psi f_0(980)$, $\psi^{\prime}f_0(980)^*$, the correlation function  is
\begin{eqnarray}
\Pi_{\mu\nu}(p,q)&=&i^2\int d^4xd^4y e^{ipx}e^{iqy}\langle 0|T\Big\{J_{J/\psi,\mu}(x)J_{f_0}(y)J_{\nu}^{\dagger}(0)\Big\}|0\rangle\, ,
\end{eqnarray}
where
\begin{eqnarray}
J_{J/\psi,\mu}(x)&=&\bar{c}(x)\gamma_\mu c(x)\, ,\nonumber\\
J_{f_0}(y)&=&\bar{s}(y)  s(y)\, ,\nonumber\\
J_\nu(0)&=&\frac{\varepsilon^{ijk}\varepsilon^{imn}}{\sqrt{2}}\Big[s^{Tj}(0)C c^k(0) \bar{s}^m(0)\gamma_\nu C \bar{c}^{Tn}(0)-s^{Tj}(0)C\gamma_\nu c^k(0)\bar{s}^m(0)C \bar{c}^{Tn}(0) \Big] \, .
\end{eqnarray}

For the two-body strong decay $Y(4660)\to \eta_c\, \phi(1020)$, the correlation function   is
\begin{eqnarray}
\Pi_{\mu\nu}(p,q)&=&i^2\int d^4xd^4y e^{ipx}e^{iqy}\langle 0|T\Big\{J_{\eta_c}(x)J_{\phi,\mu}(y)J_{\nu}^{\dagger}(0)\Big\}|0\rangle\, ,
\end{eqnarray}
where
\begin{eqnarray}
J_{\eta_c}(x)&=&\bar{c}(x)i\gamma_5 c(x)\, ,\nonumber\\
J_{\phi,\mu}(y)&=&\bar{s}(y)\gamma_\mu  s(y)\, .
\end{eqnarray}

For the two-body strong decay $Y(4660)\to \chi_{c0}\, \phi(1020)$, the correlation function  is
\begin{eqnarray}
\Pi_{\mu\nu}(p,q)&=&i^2\int d^4xd^4y e^{ipx}e^{iqy}\langle 0|T\Big\{J_{\chi_{c0}}(x)J_{\phi,\mu}(y)J_{\nu}^{\dagger}(0)\Big\}|0\rangle\, ,
\end{eqnarray}
where
\begin{eqnarray}
J_{\chi_{c0}}(x)&=&\bar{c}(x) c(x)\, .
\end{eqnarray}

For the two-body strong decay $Y(4660)\to D_s\,\bar{D}_s $, the correlation function   is
\begin{eqnarray}
\Pi_{\nu}(p,q)&=&i^2\int d^4xd^4y e^{ipx}e^{iqy}\langle 0|T\Big\{J^{\dagger}_{D_s}(x)J_{D_s}(y)J_{\nu}^{\dagger}(0)\Big\}|0\rangle\, ,
\end{eqnarray}
where
\begin{eqnarray}
J_{D_s}(y)&=&\bar{s}(y)i\gamma_5 c(y)\, .
\end{eqnarray}

For the two-body strong decay $Y(4660)\to D^*_s\,\bar{D}^*_s $, the correlation function   is
\begin{eqnarray}
\Pi_{\alpha\beta\nu}(p,q)&=&i^2\int d^4xd^4y e^{ipx}e^{iqy}\langle 0|T\Big\{J^{\dagger}_{D^*_s,\alpha}(x)J_{D^*_s,\beta}(y)J_{\nu}^{\dagger}(0)\Big\}|0\rangle\, ,
\end{eqnarray}
where
\begin{eqnarray}
J_{D_s^*,\beta}(y)&=&\bar{s}(y)\gamma_\beta c(y)\, .
\end{eqnarray}

For the two-body strong decay $Y(4660)\to D_s\,\bar{D}^*_s $, the correlation function   is
\begin{eqnarray}
\Pi_{\mu\nu}(p,q)&=&i^2\int d^4xd^4y e^{ipx}e^{iqy}\langle 0|T\Big\{J^{\dagger}_{D^*_s,\mu}(x)J_{D_s}(y)J_{\nu}^{\dagger}(0)\Big\}|0\rangle\, .
\end{eqnarray}

For the two-body strong decay $Y(4660)\to J/\psi\,\phi(1020) $, the correlation function   is
\begin{eqnarray}
\Pi_{\alpha\beta\nu}(p,q)&=&i^2\int d^4xd^4y e^{ipx}e^{iqy}\langle 0|T\Big\{J_{J/\psi,\alpha}(x)J_{\phi,\beta}(y)J_{\nu}^{\dagger}(0)\Big\}|0\rangle\, .
\end{eqnarray}

At the phenomenological side,  we insert  a complete set of intermediate hadronic states with the same quantum numbers as the current operators  into the three-point
correlation functions  and  isolate the ground state contributions to obtain the  hadron representation \cite{SVZ79,Reinders85}.

For the decays  $Y(4660)\to J/\psi f_0(980)$, $\psi^{\prime}f_0(980)^*$, the correlation function can be written as
\begin{eqnarray}
\Pi_{\mu\nu}(p,q)&=& \frac{f_{J/\psi}m_{J/\psi}f_{f_0}m_{f_0}\,\lambda_Y\,G_{YJ/\psi f_0}}{\left(p^{\prime2}-m_Y^2\right)\left(p^2-m_{J/\psi}^2 \right)\left(q^2-m_{f_0}^2\right)}\left(-g_{\mu\alpha}+\frac{p_{\mu}p_{\alpha}}{p^2}\right)\left(-g_\nu{}^\alpha+\frac{p^\prime_{\nu}p^{\prime\alpha}}{p^{\prime2}}\right)+\nonumber\\
&&\frac{f_{\psi^\prime}m_{\psi^\prime}f_{f_0}m_{f_0}\,\lambda_Y\,G_{Y\psi^\prime f_0}}{\left(p^{\prime2}-m_Y^2\right)\left(p^2-m_{\psi^\prime}^2 \right)\left(q^2-m_{f_0}^2\right)}\left(-g_{\mu\alpha}+\frac{p_{\mu}p_{\alpha}}{p^2}\right)\left(-g_\nu{}^\alpha+\frac{p^\prime_{\nu}p^{\prime\alpha}}{p^{\prime2}}\right)+\cdots\nonumber\\
&=&\Pi(p^{\prime2},p^2,q^2)\,g_{\mu\nu}+\cdots \, .
\end{eqnarray}

For the decay $Y(4660)\to \eta_c\, \phi(1020)$, the correlation function can be written as
\begin{eqnarray}
\Pi_{\mu\nu}(p,q)&=& \frac{f_{\eta_c}m_{\eta_c}^2}{2m_c}\frac{f_{\phi}m_{\phi}\,\lambda_Y\,G_{Y\eta_c\phi}\,\varepsilon_{\alpha\beta\rho\sigma}q^\alpha p^{\prime\rho}}{\left(p^{\prime2}-m_Y^2\right)\left(p^2-m_{\eta_c}^2 \right)\left(q^2-m_{\phi}^2\right)}\left(-g_{\mu}{}^{\beta}+\frac{q_{\mu}q^{\beta}}{q^2}\right)\left(-g_\nu{}^\sigma+\frac{p^\prime_{\nu}p^{\prime\sigma}}{p^{\prime2}}\right)+\cdots\nonumber\\
&=&\Pi(p^{\prime2},p^2,q^2)\,\varepsilon_{\mu\nu\alpha\beta}p^{\alpha}q^{\beta}+\cdots \, .
\end{eqnarray}

For the decay $Y(4660)\to \chi_{c0}\, \phi(1020)$, the correlation function can be written as
\begin{eqnarray}
\Pi_{\mu\nu}(p,q)&=& \frac{f_{\chi_{c0}}m_{\chi_{c0}}f_{\phi}m_{\phi}\,\lambda_Y\,G_{Y\chi_{c0} \phi}}{\left(p^{\prime2}-m_Y^2\right)\left(p^2-m_{\chi_{c0}}^2 \right)\left(q^2-m_{\phi}^2\right)}\left(-g_{\mu\alpha}+\frac{q_{\mu}q_{\alpha}}{q^2}\right)\left(-g_\nu{}^\alpha+\frac{p^\prime_{\nu}p^{\prime\alpha}}{p^{\prime2}}\right)+\cdots\nonumber\\
&=&\Pi(p^{\prime2},p^2,q^2)\,g_{\mu\nu}+\cdots \, .
\end{eqnarray}

For the decay $Y(4660)\to D_s\,\bar{D}_s $, the correlation function can be written as
\begin{eqnarray}
\Pi_{\nu}(p,q)&=& \frac{f_{D_s}^2m_{D_s}^4}{(m_c+m_s)^2}\frac{ \lambda_Y\,G_{YD_s\bar{D}_s}}{\left(p^{\prime2}-m_Y^2\right)\left(p^2-m_{D_s}^2 \right)\left(q^2-m_{D_s}^2\right)}\left(p-q\right)^\alpha\left(-g_{\alpha\nu}+\frac{p^{\prime}_{\alpha}p^\prime_{\nu}}{p^{\prime2}}\right)+\cdots\nonumber\\
&=&\Pi(p^{\prime2},p^2,q^2)\,\left( -p_{\nu}\right)+\cdots \, .
\end{eqnarray}

For the decay $Y(4660)\to D^*_s\,\bar{D}^*_s $, the correlation function can be written as
\begin{eqnarray}
\Pi_{\alpha\beta\nu}(p,q)&=& \frac{f^2_{D^*_s}m^2_{D^*_s}\,\lambda_Y\,G_{YD^*_s\bar{D}_s^*}}{\left(p^{\prime2}-m_Y^2\right)\left(p^2-m_{D_s^*}^2 \right)\left(q^2-m_{D_s^*}^2\right)}\left(p-q\right)^\sigma \left(-g_{\nu\sigma}+\frac{p^{\prime}_{\nu}p^\prime_{\sigma}}{p^{\prime2}}\right) \left(-g_{\alpha\rho} +\frac{p_{\alpha}p_{\rho}}{p^2}\right)\nonumber\\
&&\left(-g_{\beta}{}^\rho+\frac{q_{\beta}q^{\rho}}{q^2}\right)+\cdots\nonumber\\
&=&\Pi(p^{\prime2},p^2,q^2)\,\left( -g_{\alpha\beta}p_\nu\right)+\cdots \, .
\end{eqnarray}

For the decay $Y(4660)\to D_s\,\bar{D}^*_s $, the correlation function can be written as
\begin{eqnarray}
\Pi_{\mu\nu}(p,q)&=& \frac{f_{D_s}m_{D_s}^2}{m_c+m_s}\frac{f_{D^*_s}m_{D^*_s}\,\lambda_Y\,G_{YD_s\bar{D}^*_s}\,\varepsilon_{\alpha\beta\rho\sigma}p^\alpha p^{\prime\rho}}{\left(p^{\prime2}-m_Y^2\right)\left(p^2-m_{D_s^*}^2 \right)\left(q^2-m_{D_s}^2\right)}\left(-g_{\mu}{}^{\beta}+\frac{p_{\mu}p^{\beta}}{p^2}\right)\left(-g_{\nu}{}^{\sigma}+\frac{p^\prime_{\nu}p^{\prime\sigma}}{p^{\prime2}}\right)+\cdots\nonumber\\
&=&\Pi(p^{\prime2},p^2,q^2)\,\left(-\varepsilon_{\mu\nu\alpha\beta}p^{\alpha}q^{\beta}\right)+\cdots \, .
\end{eqnarray}

For the decay $Y(4660)\to J/\psi\,\phi(1020)$, the correlation function can be written as
\begin{eqnarray}
\Pi_{\alpha\beta\nu}(p,q)&=& \frac{f_{J/\psi}m_{J/\psi}f_{\phi}m_{\phi}\,\lambda_Y\,G_{YJ/\psi\phi}}{\left(p^{\prime2}-m_Y^2\right)\left(p^2-m_{J/\psi}^2 \right)\left(q^2-m_{\phi}^2\right)}\left(p-q\right)^\sigma \left(-g_{\nu\sigma}+\frac{p^{\prime}_{\nu}p^\prime_{\sigma}}{p^{\prime2}}\right) \left(-g_{\alpha\rho} +\frac{p_{\alpha}p_{\rho}}{p^2}\right)\nonumber\\
&&\left(-g_{\beta}{}^\rho+\frac{q_{\beta}q^{\rho}}{q^2}\right)+\cdots\nonumber\\
&=&\Pi(p^{\prime2},p^2,q^2)\,\left( -g_{\alpha\beta}p_\nu\right)+\cdots \, .
\end{eqnarray}
In calculations, we observe that the hadronic coupling constant $G_{YJ/\psi\phi}$ is zero at the leading order approximation, and we will neglect the process  $Y(4660)\to J/\psi\,\phi(1020)$.

In Eqs.(31-37), we have used the following definitions for the decay constants and hadronic coupling constants,
\begin{eqnarray}
\langle 0|J_{J/\psi,\mu}(0)|J/\psi(p)\rangle&=&f_{J/\psi}m_{J/\psi}\xi^{J/\psi}_\mu\, ,\nonumber\\
\langle 0|J_{\psi^\prime,\mu}(0)|\psi^\prime(p)\rangle&=&f_{\psi^\prime}m_{\psi^\prime}\xi^{\psi^\prime}_\mu\, ,\nonumber\\
\langle 0|J_{f_0}(0)|f_0(p)\rangle&=&f_{f_0}m_{f_0}\, ,\nonumber\\
\langle 0|J_{\eta_c}(0)|\eta_c(p)\rangle&=&\frac{f_{\eta_c}m_{\eta_c}^2}{2m_c}\, ,\nonumber\\
\langle 0|J_{\phi,\mu}(0)|\phi(p)\rangle&=&f_{\phi}m_{\phi}\xi^{\phi}_\mu\, ,\nonumber\\
\langle 0|J_{\chi_{c0}}(0)|\chi_{c0}(p)\rangle&=&f_{\chi_{c0}}m_{\chi_{c0}}\, ,\nonumber\\
\langle 0|J_{D_s}(0)|D_{s}(p)\rangle&=&\frac{f_{D_s}m_{D_s}^2}{m_c+m_s}\, ,\nonumber\\
\langle 0|J_{D_s^*,\mu}(0)|D_{s}^*(p)\rangle&=&f_{D_s^*}m_{D_s^*}\xi^{D^*_s}_\mu\, ,\nonumber\\
\langle 0|J_{\mu}(0)|Y(p)\rangle&=&\lambda_{Y}\xi^{Y}_\mu\, ,
\end{eqnarray}
\begin{eqnarray}
\langle J/\psi(p)f_0(q)|X(p^\prime)\rangle &=& i\,\xi^{*\alpha}_{J/\psi}\xi_\alpha^{Y}\, G_{YJ/\psi f_0}\, ,\nonumber\\
\langle \psi^\prime(p)f_0(q)|X(p^\prime)\rangle &=& i\,\xi^{*\alpha}_{\psi^\prime}\xi_\alpha^{Y}\, G_{Y\psi^\prime f_0}\, ,\nonumber\\
\langle \eta_c(p)\phi(q)|X(p^\prime)\rangle &=& i\,\varepsilon^{\alpha\beta\rho\sigma}\,q_\alpha \xi^{\phi*}_{\beta}p^{\prime}_\rho\xi_\sigma^{Y}\, G_{Y\eta_c \phi}\, ,\nonumber\\
\langle \chi_{c0}(p)\phi(q)|X(p^\prime)\rangle &=& i\,\xi^{*\alpha}_{\phi}\xi_\alpha^{Y}\, G_{Y\chi_{c0} \phi}\, ,\nonumber\\
\langle \bar{D}_s(p)D_s(q)|X(p^\prime)\rangle &=& i\,(p-q)^\alpha\xi_\alpha^{Y}\, G_{YD_s\bar{D}_s}\, ,\nonumber\\
\langle \bar{D}^*_s(p)D_s^*(q)|X(p^\prime)\rangle &=& i\,(p-q)^\alpha\xi_\alpha^{Y}\xi^{\bar{D}_s^* *}_\beta\xi^{D_s^* *\beta}\, G_{YD_s^*\bar{D}_s^*}\, ,\nonumber\\
\langle \bar{D}^*_s(p)D_s(q)|X(p^\prime)\rangle &=& i\,\varepsilon^{\alpha\beta\rho\sigma}\,p_\alpha \xi^{\bar{D}_s^* *}_{\beta}p^{\prime}_\rho\xi_\sigma^{Y}\, G_{YD_s\bar{D}_s^* }\, , \nonumber\\
\langle J/\psi(p)\phi(q)|X(p^\prime)\rangle &=& i\,(p-q)^\alpha\xi_\alpha^{Y}\xi^{J/\psi *}_\beta\xi^{\phi *\beta}\, G_{YJ/\psi \phi}\, ,
\end{eqnarray}
where the $\xi^{J/\psi}_\mu$,  $\xi^{\psi^\prime}_\mu$, $\xi^{\phi}_\mu$, $\xi^{D^*_s}_\mu$, $\xi^{Y}_\mu$ are the polarization vectors, the $G_{YJ/\psi f_0}$, $G_{Y\psi^{\prime} f_0}$, $ G_{Y\eta_c \phi}$,
$G_{Y\chi_{c0} \phi}$, $G_{YD_s\bar{D}_s}$, $G_{YD_s^*\bar{D}_s^*}$, $G_{YD_s\bar{D}_s^*}$, $G_{YJ/\psi\phi}$   are the hadronic coupling constants.

We study the components  $\Pi(p^{\prime2},p^2,q^2)$ of the correlation functions, and  carry out the operator product expansion up to the vacuum condensates of dimension 5 and neglect the tiny contributions of the gluon condensate. Then we obtain the QCD spectral densities through dispersion relation  and use Eqs.(17-18) to obtain the QCD sum rules for the hadronic coupling constants,

\begin{eqnarray}
 &&\frac{f_{J/\psi}m_{J/\psi}f_{f_0}m_{f_0}\,\lambda_Y\,G_{YJ/\psi f_0}}{m_Y^2-m_{J/\psi}^2 } \left[\exp\left(-\frac{m_{J/\psi}^2}{T_1^2}\right) -\exp\left(-\frac{m_Y^2}{T_1^2}\right) \right]\exp\left( -\frac{m_{f_0}^2}{T_2^2}\right)\nonumber\\
 &&+\left(C_{Y^{\prime}J/\psi}+C_{Y^{\prime}f_0} \right)\exp\left(-\frac{m_{J/\psi}^2}{T_1^2} -\frac{m_{f_0}^2}{T_2^2}\right)\nonumber\\
&=&-\frac{1}{32\sqrt{2}\pi^4}\int_{4m_c^2}^{s^0_{J/\psi}} ds \int_0^{s^0_{f_0}} du us
\sqrt{1-\frac{4m_c^2}{s}}\left(1+\frac{2m_c^2}{s}\right)\exp\left(-\frac{s}{T_1^2}-\frac{u}{T_2^2}\right)\nonumber\\
&&-\frac{m_s\langle\bar{s}s\rangle}{4\sqrt{2}\pi^2}\int_{4m_c^2}^{s^0_{J/\psi}} ds s
\sqrt{1-\frac{4m_c^2}{s}}\left(1+\frac{2m_c^2}{s}\right)\exp\left(-\frac{s}{T_1^2}\right)\nonumber\\
&&-\frac{m_s\langle\bar{s}g_{s}\sigma Gs\rangle}{12\sqrt{2}\pi^2T_2^2}\int_{4m_c^2}^{s^0_{J/\psi}} ds
\sqrt{1-\frac{4m_c^2}{s}}\left(s+2m_c^2\right)\exp\left(-\frac{s}{T_1^2}\right)\nonumber\\
&&+\frac{m_s\langle\bar{s}g_{s}\sigma Gs\rangle}{48\sqrt{2}\pi^2}\int_{4m_c^2}^{s^0_{J/\psi}} ds
\frac{s-12m_c^2}{\sqrt{s\left(s-4m_c^2\right)}}\exp\left(-\frac{s}{T_1^2}\right) \, ,
\end{eqnarray}

\begin{eqnarray}
&& \frac{f_{\eta_c}m_{\eta_c}^2}{2m_c}\frac{f_{\phi}m_{\phi}\,\lambda_Y\,G_{Y\eta_c\phi}\,}{m_Y^2-m_{\eta_c}^2}\left[\exp\left(-\frac{m_{\eta_c}^2}{T_1^2}\right) -\exp\left(-\frac{m_Y^2}{T_1^2}\right) \right]\exp\left( -\frac{m_{\phi}^2}{T_2^2}\right)\nonumber\\
 &&+\left(C_{Y^{\prime}\eta_c}+C_{Y^{\prime}\phi} \right)\exp\left(-\frac{m_{\eta_c}^2}{T_1^2} -\frac{m_{\phi}^2}{T_2^2}\right)\nonumber\\
&=&\frac{3m_s m_c}{16\sqrt{2}\pi^4}\int_{4m_c^2}^{s^0_{\eta_{c}}} ds \int_0^{s^0_\phi} du
\sqrt{1-\frac{4m_c^2}{s}}\exp\left(-\frac{s}{T_1^2}-\frac{u}{T_2^2}\right)\nonumber\\
&&-\frac{m_c\langle\bar{s}s\rangle}{2\sqrt{2}\pi^2}\int_{4m_c^2}^{s^0_{\eta_{c}}} ds
\sqrt{1-\frac{4m_c^2}{s}}\exp\left(-\frac{s}{T_1^2}\right)\nonumber\\
&&+\frac{m_c\langle\bar{s}g_{s}\sigma Gs\rangle}{6\sqrt{2}\pi^2T_2^2}\int_{4m_c^2}^{s^0_{\eta_{c}}} ds
\sqrt{1-\frac{4m_c^2}{s}}\exp\left(-\frac{s}{T_1^2}\right)\nonumber\\
&&-\frac{m_c\langle\bar{s}g_{s}\sigma Gs\rangle}{24\sqrt{2}\pi^2}\int_{4m_c^2}^{s^0_{\eta_{c}}} ds
\frac{1}{\sqrt{s\left(s-4m_c^2\right)}}\exp\left(-\frac{s}{T_1^2}\right) \, ,
\end{eqnarray}

\begin{eqnarray}
&& \frac{f_{\chi_{c0}}m_{\chi_{c0}}f_{\phi}m_{\phi}\,\lambda_Y\,G_{Y\chi_{c0} \phi}}{m_Y^2-m_{\chi_{c0}}^2}\left[\exp\left(-\frac{m_{\chi_{c0}}^2}{T_1^2}\right) -\exp\left(-\frac{m_Y^2}{T_1^2}\right) \right]\exp\left( -\frac{m_{\phi}^2}{T_2^2}\right)\nonumber\\
 &&+\left(C_{Y^{\prime}\chi_{c0}}+C_{Y^{\prime}\phi} \right)\exp\left(-\frac{m_{\chi_{c0}}^2}{T_1^2} -\frac{m_{\phi}^2}{T_2^2}\right)\nonumber\\
&=&\frac{1}{32\sqrt{2}\pi^4}\int_{4m_c^2}^{s^0_{\chi_{c0}}} ds \int_0^{s^0_\phi} duus
\sqrt{1-\frac{4m_c^2}{s}}\left(1-\frac{4m_c^2}{s}\right)\exp\left(-\frac{s}{T_1^2}-\frac{u}{T_2^2}\right) \nonumber\\
&&-\frac{m_s\langle\bar{s}s\rangle}{4\sqrt{2}\pi^2}\int_{4m_c^2}^{s^0_{\chi_{c0}}} ds s
\sqrt{1-\frac{4m_c^2}{s}}\left(1-\frac{4m_c^2}{s}\right)\exp\left(-\frac{s}{T_1^2}\right) \nonumber\\
&&+\frac{m_s\langle\bar{s}g_{s}\sigma Gs\rangle}{48\sqrt{2}\pi^2T_2^2}\int_{4m_c^2}^{s^0_{\chi_{c0}}} ds s
\sqrt{1-\frac{4m_c^2}{s}}\left(1-\frac{4m_c^2}{s}\right)\exp\left(-\frac{s}{T_1^2}\right) \nonumber\\
&&-\frac{m_s\langle\bar{s}g_{s}\sigma Gs\rangle}{24\sqrt{2}\pi^2}\int_{4m_c^2}^{s^0_{\chi_{c0}}} ds
\frac{s-6m_c^2}{\sqrt{s\left(s-4m_c^2\right)}}\exp\left(-\frac{s}{T_1^2}\right) \, ,
\end{eqnarray}

\begin{eqnarray}
&& \frac{f_{D_s}^2m_{D_s}^4}{(m_c+m_s)^2}\frac{ \lambda_Y\,G_{YD_s\bar{D}_s}}{4\left(\widetilde{m}_Y^2-m_{D_s}^2\right)}\left[\exp\left(-\frac{m_{D_s}^2}{T_1^2}\right) -\exp\left(-\frac{\widetilde{m}_Y^2}{T_1^2}\right) \right]\exp\left( -\frac{m_{D_s}^2}{T_2^2}\right)\nonumber\\
 &&+\left(C_{Y^{\prime}\bar{D}_s}+C_{Y^{\prime}D_s} \right)\exp\left(-\frac{m_{D_s}^2}{T_1^2} -\frac{m_{D_s}^2}{T_2^2}\right)\nonumber\\
&=&\frac{3m_c}{64\sqrt{2}\pi^4}\int_{m_c^2}^{s^0_{D_s}} ds \int_{m_c^2}^{s^0_{D_s}} du u
\left(1-\frac{m_c^2}{s}\right)^2\left(1-\frac{m_c^2}{u}\right)^2\exp\left(-\frac{s}{T_1^2}-\frac{u}{T_2^2}\right) \nonumber\\
&&+\frac{3m_s}{64\sqrt{2}\pi^4}\int_{m_c^2}^{s^0_{D_s}} ds \int_{m_c^2}^{s^0_{D_s}} du u
\left(1-\frac{m_c^2}{s}\right)\left(1-\frac{m_c^2}{u}\right)\left(1+\frac{m_c^2}{s}+\frac{m_c^2}{u}-\frac{3m_c^4}{us}\right)
\exp\left(-\frac{s}{T_1^2}-\frac{u}{T_2^2}\right)\nonumber\\
&&-\frac{\langle\bar{s}s\rangle}{8\sqrt{2}\pi^2}\int_{m_c^2}^{s^0_{D_s}} du
\left[u\left(1-\frac{m_c^2}{u}\right)^2+2m_s m_c\left(1-\frac{m_c^2}{u}\right)\right]\exp\left(-\frac{m_c^2}{T_1^2}-\frac{u}{T_2^2}\right) \nonumber\\
&&-\frac{\langle\bar{s}s\rangle}{8\sqrt{2}\pi^2}\int_{m_c^2}^{s^0_{D_s}} ds
\left[m_c^2\left(1-\frac{m_c^2}{s}\right)^2+m_s m_c\left(1-\frac{m_c^4}{s^2}\right)\right]\exp\left(-\frac{s}{T_1^2}-\frac{m_c^2}{T_2^2}\right) \nonumber\\
&&+\frac{m_s m_c\langle\bar{s}s\rangle}{16\sqrt{2}\pi^2T_1^2}\int_{m_c^2}^{s^0_{D_s}} du u
\left(1-\frac{m_c^2}{u}\right)^2\exp\left(-\frac{m_c^2}{T_1^2}-\frac{u}{T_2^2}\right) \nonumber\\
&&+\frac{m_s m_c\langle\bar{s}s\rangle}{16\sqrt{2}\pi^2}\int_{m_c^2}^{s^0_{D_s}} ds
\left(1-\frac{m_c^2}{s}\right)^2\left(1+\frac{m_c^2}{T_2^2}\right)\exp\left(-\frac{s}{T_1^2}-\frac{m_c^2}{T_2^2}\right) \nonumber\\
&&+\frac{m_c^2\langle\bar{s}g_{s}\sigma Gs\rangle}{32\sqrt{2}\pi^2T_1^4}\int_{m_c^2}^{s^0_{D_s}} du
\left[u\left(1-\frac{m_c^2}{u}\right)^2+2m_s m_c\left(1-\frac{m_c^2}{u}\right)\right]\exp\left(-\frac{m_c^2}{T_1^2}-\frac{u}{T_2^2}\right) \nonumber\\
&&-\frac{\langle\bar{s}g_{s}\sigma Gs\rangle}{16\sqrt{2}\pi^2T_2^2}\int_{m_c^2}^{s^0_{D_s}} ds
\left[m_c^2\left(1-\frac{m_c^2}{s}\right)^2+m_s m_c\left(1-\frac{m_c^4}{s^2}\right)\right]\left(1-\frac{m_c^2}{2T_2^2}\right)
\exp\left(-\frac{s}{T_1^2}-\frac{m_c^2}{T_2^2}\right) \nonumber\\
&&-\frac{m_s m_c^3\langle\bar{s}g_{s}\sigma Gs\rangle}{96\sqrt{2}\pi^2T_1^6}\int_{m_c^2}^{s^0_{D_s}} du u
\left(1-\frac{m_c^2}{u}\right)^2\exp\left(-\frac{m_c^2}{T_1^2}-\frac{u}{T_2^2}\right) \nonumber\\
&&-\frac{m_s m_c^5\langle\bar{s}g_{s}\sigma Gs\rangle}{96\sqrt{2}\pi^2T_2^6}\int_{m_c^2}^{s^0_{D_s}} ds
\left(1-\frac{m_c^2}{s}\right)^2\exp\left(-\frac{s}{T_1^2}-\frac{m_c^2}{T_2^2}\right) \nonumber\\
&&-\frac{\langle\bar{s}g_{s}\sigma Gs\rangle}{192\sqrt{2}\pi^2}\int_{m_c^2}^{s^0_{D_s}} du
\left[\left(1-\frac{m_c^2}{u}\right)\left(3-\frac{m_c^2}{u}\right)+\frac{6m_s m_c}{u}\right]
\exp\left(-\frac{m_c^2}{T_1^2}-\frac{u}{T_2^2}\right) \nonumber\\
&&-\frac{\langle\bar{s}g_{s}\sigma Gs\rangle}{96\sqrt{2}\pi^2}\int_{m_c^2}^{s^0_{D_s}} ds
\left[\frac{m_c^2}{s}\left(1-\frac{m_c^2}{s}\right)\left(6-\frac{m_c^2}{s}\right)
-\frac{3m_s m_c^3}{s^2}\right]\exp\left(-\frac{s}{T_1^2}-\frac{m_c^2}{T_2^2}\right) \nonumber\\
&&-\frac{\langle\bar{s}g_{s}\sigma Gs\rangle}{192\sqrt{2}\pi^2}\int_{m_c^2}^{s^0_{D_s}} du
\left[\left(3-\frac{m_c^4}{u^2}\right)+\frac{6m_s m_c}{u}\frac{u+m_c^2}{u-m_c^2}\right]
\exp\left(-\frac{m_c^2}{T_1^2}-\frac{u}{T_2^2}\right)\nonumber\\
&&-\frac{\langle\bar{s}g_{s}\sigma Gs\rangle}{96\sqrt{2}\pi^2}\int_{m_c^2}^{s^0_{D_s}} ds
\left(\frac{m_c^6}{s^3}+\frac{3m_s m_c}{s^2}\frac{s^2+m_c^4}{s-m_c^2}\right)\exp\left(-\frac{s}{T_1^2}-\frac{m_c^2}{T_2^2}\right) \, ,
\end{eqnarray}

\begin{eqnarray}
&& \frac{f^2_{D^*_s}m^2_{D^*_s}\,\lambda_Y\,G_{YD^*_s\bar{D}_s^*}}{4\left(\widetilde{m}_Y^2-m_{D_s^*}^2\right)}\left[\exp\left(-\frac{m_{D_s^*}^2}{T_1^2}\right) -\exp\left(-\frac{\widetilde{m}_Y^2}{T_1^2}\right) \right]\exp\left( -\frac{m_{D_s^*}^2}{T_2^2}\right)\nonumber\\
 &&+\left(C_{Y^{\prime}\bar{D}_s^*}+C_{Y^{\prime}D_s^*} \right)\exp\left(-\frac{m_{D_s^*}^2}{T_1^2} -\frac{m_{D_s^*}^2}{T_2^2}\right)\nonumber\\
&=&\frac{m_c}{64\sqrt{2}\pi^4}\int_{m_c^2}^{s^0_{ D_s^*}} ds \int_{m_c^2}^{s^0_{D_s^*}} du
\left(2u+m_c^2\right)\left(1-\frac{m_c^2}{s}\right)^2\left(1-\frac{m_c^2}{u}\right)^2\exp\left(-\frac{s}{T_1^2}-\frac{u}{T_2^2}\right) \nonumber\\
&&+\frac{m_s}{64\sqrt{2}\pi^4}\int_{m_c^2}^{s^0_{D_s^*}} ds \int_{m_c^2}^{s^0_{D_s^*}} du u\left(1-\frac{m_c^2}{s}\right)\left(1-\frac{m_c^2}{u}\right) \nonumber\\
&&\left(2+\frac{2m_c^2}{s}+\frac{5m_c^2}{u}-\frac{m_c^4}{u^2}-\frac{7m_c^4}{us}-\frac{m_c^6}{us^2}\right)
\exp\left(-\frac{s}{T_1^2}-\frac{u}{T_2^2}\right) \nonumber\\
&&-\frac{\langle\bar{s}s\rangle}{24\sqrt{2}\pi^2}\int_{m_c^2}^{s^0_{D_s^*}} du\left[\left(2u+m_c^2\right)\left(1-\frac{m_c^2}{u}\right)^2+6m_s m_c\left(1-\frac{m_c^2}{u}\right)\right]\exp\left(-\frac{m_c^2}{T_1^2}-\frac{u}{T_2^2}\right) \nonumber\\
&&-\frac{\langle\bar{s}s\rangle}{8\sqrt{2}\pi^2} \int_{m_c^2}^{s^0_{D_s^*}} ds\left[m_c^2\left(1-\frac{m_c^2}{s}\right)^2+m_s m_c\left(1-\frac{m_c^4}{s^2}\right)\right]
\exp\left(-\frac{s}{T_1^2}-\frac{m_c^2}{T_2^2}\right) \nonumber\\
&&+\frac{m_s m_c\langle\bar{s}s\rangle}{48\sqrt{2}\pi^2T_1^2}\int_{m_c^2}^{s^0_{D_s^*}} du
\left(2u+m_c^2\right)\left(1-\frac{m_c^2}{u}\right)^2 \exp\left(-\frac{m_c^2}{T_1^2}-\frac{u}{T_2^2}\right) \nonumber\\
&&+\frac{m_s m_c^3\langle\bar{s}s\rangle}{16\sqrt{2}\pi^2T_2^2} \int_{m_c^2}^{s^0_{D_s^*}} ds
\left(1-\frac{m_c^2}{s}\right)^2\exp\left(-\frac{s}{T_1^2}-\frac{m_c^2}{T_2^2}\right) \nonumber\\
&&+\frac{\langle\bar{s}g_{s}\sigma Gs\rangle}{288\sqrt{2}\pi^2T_1^2}\int_{m_c^2}^{s^0_{D_s^*}} du
\left[\left(2u+m_c^2\right)\left(1-\frac{m_c^2}{u}\right)^2+6m_s m_c\left(1-\frac{m_c^2}{u}\right)\right]
\left(4+\frac{3m_c^2}{T_1^2}\right)\exp\left(-\frac{m_c^2}{T_1^2}-\frac{u}{T_2^2}\right) \nonumber\\
&&+\frac{m_c^2\langle\bar{s}g_{s}\sigma Gs\rangle}{32\sqrt{2}\pi^2T_2^4} \int_{m_c^2}^{s^0_{D_s^*}} ds
\left[m_c^2\left(1-\frac{m_c^2}{s}\right)^2+m_s m_c\left(1-\frac{m_c^4}{s^2}\right)\right]
\exp\left(-\frac{s}{T_1^2}-\frac{m_c^2}{T_2^2}\right) \nonumber\\
&&-\frac{m_s m_c^3\langle\bar{s}g_{s}\sigma Gs\rangle}{288\sqrt{2}\pi^2T_1^6}\int_{m_c^2}^{s^0_{D_s^*}} du
\left(2u+m_c^2\right)\left(1-\frac{m_c^2}{u}\right)^2\exp\left(-\frac{m_c^2}{T_1^2}-\frac{u}{T_2^2}\right) \nonumber\\
&&+\frac{m_s m_c\langle\bar{s}g_{s}\sigma Gs\rangle}{96\sqrt{2}\pi^2T_2^2} \int_{m_c^2}^{s^0_{D_s^*}} ds
\left(1-\frac{m_c^2}{s}\right)^2\left(1+\frac{m_c^2}{T_2^2}-\frac{m_c^4}{T_2^4}\right)\exp\left(-\frac{s}{T_1^2}-\frac{m_c^2}{T_2^2}\right) \nonumber\\
&&-\frac{\langle\bar{s}g_{s}\sigma Gs\rangle}{96\sqrt{2}\pi^2}\int_{m_c^2}^{s^0_{D_s^*}} du
\left[\left(1-\frac{m_c^2}{u}\right)+\frac{3m_s m_c}{u}\right]\exp\left(-\frac{m_c^2}{T_1^2}-\frac{u}{T_2^2}\right) \nonumber\\
&&-\frac{\langle\bar{s}g_{s}\sigma Gs\rangle}{96\sqrt{2}\pi^2} \int_{m_c^2}^{s^0_{D_s^*}} ds
\left[\frac{m_c^2}{s}\left(1-\frac{m_c^2}{s}\right)\left(6-\frac{m_c^2}{s}\right)-\frac{3m_s m_c^3}{s^2}\right]
\exp\left(-\frac{s}{T_1^2}-\frac{m_c^2}{T_2^2}\right) \nonumber\\
&&+\frac{\langle\bar{s}g_{s}\sigma Gs\rangle}{96\sqrt{2}\pi^2}\int_{m_c^2}^{s^0_{D_s^*}} du
\left(1+\frac{m_s m_c}{u}\frac{u+m_c^2}{u-m_c^2}\right)\exp\left(-\frac{m_c^2}{T_1^2}-\frac{u}{T_2^2}\right)\nonumber\\
&&+\frac{\langle\bar{s}g_{s}\sigma Gs\rangle}{96\sqrt{2}\pi^2} \int_{m_c^2}^{s^0_{D_s^*}} ds
\left(\frac{m_c^4}{s^2}+\frac{m_s m_c}{s^2}\frac{s^2+m_c^4}{s-m_c^2}\right)\exp\left(-\frac{s}{T_1^2}-\frac{m_c^2}{T_2^2}\right)\, ,
\end{eqnarray}

\begin{eqnarray}
&& \frac{f_{D_s}m_{D_s}^2}{m_c+m_s}\frac{f_{D^*_s}m_{D^*_s}\,\lambda_Y\,G_{YD_s\bar{D}^*_s} }{4\left(\widetilde{m}_Y^2-m_{D_s^*}^2\right)}\left[\exp\left(-\frac{m_{D_s^*}^2}{T_1^2}\right) -\exp\left(-\frac{\widetilde{m}_Y^2}{T_1^2}\right) \right]\exp\left( -\frac{m_{D_s}^2}{T_2^2}\right)\nonumber\\
 &&+\left(C_{Y^{\prime}\bar{D}_s^*}+C_{Y^{\prime}D_s} \right)\exp\left(-\frac{m_{D_s^*}^2}{T_1^2} -\frac{m_{D_s}^2}{T_2^2}\right)\nonumber\\
&=&-\frac{\langle\bar{s}g_{s}\sigma Gs\rangle}{32\sqrt{2}\pi^2}\int_{m_c^2}^{s^0_{D_s}} du
\left[\frac{m_c}{u}\left(1-\frac{m_c^2}{u}\right)+\frac{2m_s m_c^2}{3u^2}\right]\exp\left(-\frac{m_c^2}{T_1^2}-\frac{u}{T_2^2}\right) \nonumber\\
&&+\frac{\langle\bar{s}g_{s}\sigma Gs\rangle}{32\sqrt{2}\pi^2}\int_{m_c^2}^{s^0_{D_s^*}} ds
\left[\frac{m_c}{s}\left(1-\frac{m_c^2}{s}\right)-\frac{2m_s m_c^2}{3s^2}\right]\exp\left(-\frac{s}{T_1^2}-\frac{m_c^2}{T_2^2}\right) \nonumber\\
&&-\frac{\langle\bar{s}g_{s}\sigma Gs\rangle}{48\sqrt{2}\pi^2}\int_{m_c^2}^{s^0_{D_s}} du
\left(\frac{m_c^3}{u^2}+\frac{m_s}{2u^2}\frac{u^2+m_c^4}{u-m_c^2}\right)\exp\left(-\frac{m_c^2}{T_1^2}-\frac{u}{T_2^2}\right)
\nonumber\\
&&-\frac{\langle\bar{s}g_{s}\sigma Gs\rangle}{96\sqrt{2}\pi^2}\int_{m_c^2}^{s^0_{D_s^*}} ds
\left(\frac{m_s}{s^2}\frac{s^2+m_c^4}{s-m_c^2}\right)\exp\left(-\frac{s}{T_1^2}-\frac{m_c^2}{T_2^2}\right) \, ,
\end{eqnarray}

\begin{eqnarray}
 &&\frac{f_{J/\psi}m_{J/\psi}f_{f_0}m_{f_0}\,\lambda_Y\,G_{YJ/\psi f_0}}{m_Y^2-m_{J/\psi}^2 } \left[\exp\left(-\frac{m_{J/\psi}^2}{T_1^2}\right) -\exp\left(-\frac{m_Y^2}{T_1^2}\right) \right]\exp\left( -\frac{m_{f_0}^2}{T_2^2}\right)\nonumber\\
 &&+\left(C_{Y^{\prime}J/\psi}+C_{Y^{\prime}f_0} \right)\exp\left(-\frac{m_{J/\psi}^2}{T_1^2} -\frac{m_{f_0}^2}{T_2^2}\right)\nonumber\\
 &&+\frac{f_{\psi^\prime}m_{\psi^\prime}f_{f_0}m_{f_0}\,\lambda_Y\,G_{Y\psi^\prime f_0}}{m_Y^2-m_{\psi^\prime}^2 } \left[\exp\left(-\frac{m_{\psi^\prime}^2}{T_1^2}\right) -\exp\left(-\frac{m_Y^2}{T_1^2}\right) \right]\exp\left( -\frac{m_{f_0}^2}{T_2^2}\right)\nonumber\\
 &&+\left(C_{Y^{\prime}\psi^\prime}+\widetilde{C}_{Y^{\prime}f_0} \right)\exp\left(-\frac{m_{\psi^\prime}^2}{T_1^2} -\frac{m_{f_0}^2}{T_2^2}\right)\nonumber\\
&=&-\frac{1}{32\sqrt{2}\pi^4}\int_{4m_c^2}^{s^0_{\psi^\prime}} ds \int_0^{s^0_{f_0}} du us
\sqrt{1-\frac{4m_c^2}{s}}\left(1+\frac{2m_c^2}{s}\right)\exp\left(-\frac{s}{T_1^2}-\frac{u}{T_2^2}\right)\nonumber\\
&&-\frac{m_s\langle\bar{s}s\rangle}{4\sqrt{2}\pi^2}\int_{4m_c^2}^{s^0_{\psi^\prime}} ds s
\sqrt{1-\frac{4m_c^2}{s}}\left(1+\frac{2m_c^2}{s}\right)\exp\left(-\frac{s}{T_1^2}\right)\nonumber\\
&&-\frac{m_s\langle\bar{s}g_{s}\sigma Gs\rangle}{12\sqrt{2}\pi^2T_2^2}\int_{4m_c^2}^{s^0_{\psi^\prime}} ds
\sqrt{1-\frac{4m_c^2}{s}}\left(s+2m_c^2\right)\exp\left(-\frac{s}{T_1^2}\right)\nonumber\\
&&+\frac{m_s\langle\bar{s}g_{s}\sigma Gs\rangle}{48\sqrt{2}\pi^2}\int_{4m_c^2}^{s^0_{\psi^\prime}} ds
\frac{s-12m_c^2}{\sqrt{s\left(s-4m_c^2\right)}}\exp\left(-\frac{s}{T_1^2}\right) \, ,
\end{eqnarray}
where $\widetilde{m}_Y^2=\frac{m_Y^2}{4}$.
In calculations, we observe that there appears divergence due to the endpoint $s=4m_c^2$, $s=m_c^2$ and $u=m_c^2$, we can avoid the endpoint divergence with the simple replacement  $\frac{1}{s-4m_c^2} \to \frac{1}{s-4m_c^2+4m_s^2}$, $\frac{1}{u-m_c^2} \to \frac{1}{u-m_c^2+4m_s^2}$ and $\frac{1}{s-m_c^2} \to \frac{1}{s-m_c^2+4m_s^2}$  by adding a small squared $s$-quark mass $4m_s^2$.

The hadronic parameters are taken as
$m_{J/\psi}=3.0969\,\rm{GeV}$, $m_{\phi}=1.019461\,\rm{GeV}$,
$m_{\eta_c}=2.9839\,\rm{GeV}$, $m_{f_0}=0.990\,\rm{GeV}$,
$m_{D_s}=1.969\,\rm{GeV}$, $m_{D_s^*}=2.1122\,\rm{GeV}$,
$m_{\chi_{c0}}=3.41471\,\rm{GeV}$,  $m_{\psi^\prime}=3.686097\,\rm{GeV}$, $m_{\pi^+}=0.13957\,\rm{GeV}$, $f_{\psi^\prime}=0.295 \,\rm{GeV}$, $\sqrt{s^0_{J/\psi}}=3.6\,\rm{GeV}$, $\sqrt{s^0_{\eta_c}}=3.5\,\rm{GeV}$, $\sqrt{s^0_{\psi^\prime}}=4.0\,\rm{GeV}$,
$\sqrt{s^0_{D_s}}=2.5\,\rm{GeV}$, $\sqrt{s^0_{D_s^*}}=2.6\,\rm{GeV}$, $\sqrt{s^0_{\chi_{c0}}}=3.9\,\rm{GeV}$ \cite{PDG},
$f_{J/\psi}=0.418 \,\rm{GeV}$, $f_{\eta_c}=0.387 \,\rm{GeV}$  \cite{Becirevic},
$f_{\phi}=0.253\,\rm{GeV}$, $\sqrt{s^0_{\phi}}=1.5\,\rm{GeV}$   \cite{Wang-Y4274},
$f_{f_0}=0.180\,\rm{GeV}$, $\sqrt{s^0_{f_0}}=1.3\,\rm{GeV}$   \cite{Wang-f980-decay},
$f_{D_s}=0.240\,\rm{GeV}$, $f_{D_s^*}=0.308\,\rm{GeV}$  \cite{Wang-NPA-2004,Wang-heavy-decay},
$f_{\chi_{c0}}=0.359\,\rm{GeV}$ \cite{Charmonium-PRT},
$m_Y=4.652\,\rm{GeV}$ \cite{Belle-Y4660-2014},   $\lambda_{Y}=6.72\times 10^{-2}\,\rm{GeV}^5$ \cite{WangY4360Y4660-1803}. In Ref.\cite{WangY4360Y4660-1803}, we
 obtain the values $m_Y=4.66\,\rm{GeV}$ and $\lambda_{Y}=6.74\times 10^{-2}\,\rm{GeV}^5$. In this article, we choose a slightly  smaller value $\lambda_{Y}=6.72\times 10^{-2}\,\rm{GeV}^5$, which corresponds to $m_Y=4.65\,\rm{GeV}$. For more literatures on the decay constants of the charmonium or bottomonium states, one can consult Ref.\cite{Wang-PLB-charm-decay}.

 At the QCD side, we take the vacuum condensates  to be the standard values
$\langle\bar{q}q \rangle=-(0.24\pm 0.01\, \rm{GeV})^3$,  $\langle\bar{s}s \rangle=(0.8\pm0.1)\langle\bar{q}q \rangle$,
 $\langle\bar{s}g_s\sigma G s \rangle=m_0^2\langle \bar{s}s \rangle$,
$m_0^2=(0.8 \pm 0.1)\,\rm{GeV}^2$   at the energy scale  $\mu=1\, \rm{GeV}$
\cite{SVZ79,Reinders85,ColangeloReview}, and  take the $\overline{MS}$ masses $m_{c}(m_c)=(1.275\pm0.025)\,\rm{GeV}$ and $m_s(\mu=2\,\rm{GeV})=(0.095\pm0.005)\,\rm{GeV}$
 from the Particle Data Group \cite{PDG}.
Moreover,  we take into account
the energy-scale dependence of  the quark condensate, mixed quark condensate and $\overline{MS}$ masses from the renormalization group equation,
 \begin{eqnarray}
 \langle\bar{s}s \rangle(\mu)&=&\langle\bar{s}s \rangle({\rm 1GeV})\left[\frac{\alpha_{s}({\rm 1GeV})}{\alpha_{s}(\mu)}\right]^{\frac{12}{33-2n_f}}\, , \nonumber\\
 \langle\bar{s}g_s \sigma Gs \rangle(\mu)&=&\langle\bar{s}g_s \sigma Gs \rangle({\rm 1GeV})\left[\frac{\alpha_{s}({\rm 1GeV})}{\alpha_{s}(\mu)}\right]^{\frac{2}{33-2n_f}}\, ,\nonumber\\
m_c(\mu)&=&m_c(m_c)\left[\frac{\alpha_{s}(\mu)}{\alpha_{s}(m_c)}\right]^{\frac{12}{33-2n_f}} \, ,\nonumber\\
m_s(\mu)&=&m_s({\rm 2GeV} )\left[\frac{\alpha_{s}(\mu)}{\alpha_{s}({\rm 2GeV})}\right]^{\frac{12}{33-2n_f}}\, ,\nonumber\\
\alpha_s(\mu)&=&\frac{1}{b_0t}\left[1-\frac{b_1}{b_0^2}\frac{\log t}{t} +\frac{b_1^2(\log^2{t}-\log{t}-1)+b_0b_2}{b_0^4t^2}\right]\, ,
\end{eqnarray}
  where $t=\log \frac{\mu^2}{\Lambda^2}$, $b_0=\frac{33-2n_f}{12\pi}$, $b_1=\frac{153-19n_f}{24\pi^2}$, $b_2=\frac{2857-\frac{5033}{9}n_f+\frac{325}{27}n_f^2}{128\pi^3}$,  $\Lambda=210\,\rm{MeV}$, $292\,\rm{MeV}$  and  $332\,\rm{MeV}$ for the flavors  $n_f=5$, $4$ and $3$, respectively \cite{PDG,Narison-mix}, and evolve all the input parameters to the optimal  energy scale   $\mu$  with $n_f=4$ to extract the hadronic coupling constants.

  In the QCD sum rules for the mass of the $Y(4660)$, the optimal energy scale of the QCD spectral density is $\mu=2.9\,\rm{GeV}$ \cite{WangY4360Y4660-1803}, which is determined by the energy scale formula $\mu=\sqrt{M^2_{X/Y/Z}-(2{\mathbb{M}}_c)^2}$ with the updated value  of the effective $c$-quark mass ${\mathbb{M}}_c=1.82\,\rm{GeV}$ \cite{WangEPJC-1601}. In the present QCD sum rules, if we choose the energy scale $\mu=2.9\,\rm{GeV}$, we obtain an energy scale as large as the masses of the $\eta_c$ and $J/\psi$ and much larger than the masses of the $D_s$ and $D_s^*$, it is a too large energy scale.  In this article, we take the energy scales of the QCD spectral densities to be $\mu=\frac{m_{\eta_c}}{2}=1.5\,\rm{GeV}$, which is acceptable for the mesons $D$ and $J/\psi$ \cite{WangHuangTao-3900}.
  We set the Borel parameters to be $T_1^2=T_2^2=T^2$ for simplicity.
The unknown parameters are chosen as
$C_{X^{\prime}J/\psi}+C_{X^{\prime}f_0}=-0.012\,\rm{GeV}^8 $,
$C_{X^{\prime}\eta_c}+C_{X^{\prime}\phi}=0.0016\,\rm{GeV}^6 $,
$C_{X^{\prime}\chi_{c0}}+C_{X^{\prime}\phi}=0.0135\,\rm{GeV}^8 $,
$C_{X^{\prime}D_s}+C_{X^{\prime}\bar{D}_s}=0.0038\,\rm{GeV}^7 $,
$C_{X^{\prime}D_s^*}+C_{X^{\prime}\bar{D}_s^*}=0.006\,\rm{GeV}^7 $,
$C_{X^{\prime}D_s}+C_{X^{\prime}\bar{D}_s^*}=0.001\,\rm{GeV}^6 $,
$C_{X^{\prime}\psi^\prime}+\widetilde{C}_{X^{\prime}f_0}=-0.018\,\rm{GeV}^8 $
   to obtain  platforms in the Borel windows, which are shown in Table 3 explicitly. The Borel windows $T_{max}^2-T^2_{min}=1.0\,\rm{GeV}^2$ for the charmonium decays
   and  $T_{max}^2-T^2_{min}=0.8\,\rm{GeV}^2$ for the open-charm decays, where the $T^2_{max}$ and $T^2_{min}$ denote the maximum and minimum of the Borel parameters, respectively. In the Borel widows, the  platforms are flat  enough, see the central values in Figs.1-2.

\begin{table}
\begin{center}
\begin{tabular}{|c|c|c|c|c|c|c|c|}\hline\hline
                                  &$T^2(\rm{GeV}^2)$   &$|G_{ABC}|$                              &$\Gamma(\rm{MeV})$   \\ \hline

$Y(4660)\to J/\psi f_0(980)$      &$3.2-4.2$           &$1.37^{+1.16}_{-1.06}\,\rm{GeV}$         &$3.5^{+8.5}_{-3.4}$     \\ \hline

$Y(4660)\to \eta_c \phi(1020)$    &$4.3-5.3$           &$0.98^{+0.27}_{-0.25}\,\rm{GeV}^{-1}$    &$31.6^{+19.8}_{-14.0}$     \\ \hline

$Y(4660)\to \chi_{c0}\phi(1020)$  &$3.6-4.6$           &$1.17^{+1.07}_{-0.95}\,\rm{GeV}$         &$1.7^{+4.5}_{-1.6}$     \\ \hline

$Y(4660)\to D_s \bar{D}_s$        &$1.9-2.7$           &$1.36^{+0.39}_{-0.33} $                  &$8.6^{+5.6}_{-3.7}$     \\ \hline

$Y(4660)\to D_s^* \bar{D}^*_s$    &$2.5-3.3$           &$1.57^{+0.53}_{-0.50} $                  &$22.5^{+17.8}_{-11.7}$     \\ \hline

$Y(4660)\to D_s \bar{D}^*_s$      &$2.5-3.3$           &$0.11^{+0.23}_{-0.11}\,\rm{GeV}^{-1}$    &$0.4^{+1.7}_{-0.4}$     \\ \hline

$Y(4660)\to \psi^\prime f_0(980)\to \psi^\prime \pi^+\pi^-$  &$4.4-5.4$    &$7.00^{+2.24}_{-2.20}\,\rm{GeV}$     &$5.5^{+4.1}_{-2.9}$   \\ \hline

$Y(4660)\to J/\psi \phi(1020)$   &                     &$\sim 0$                                 &$\sim 0$     \\ \hline\hline

\end{tabular}
\end{center}
\caption{ The Borel  windows, hadronic coupling constants,  partial decay widths of the $Y(4660)$. }
\end{table}

In Figs.1-2, we plot the hadronic coupling constants  $G_{YBC}$  with variations of the  Borel parameters $T^2$ at much larger intervals than the Borel windows. From the figures, we can see that there appear platforms in the Borel windows indeed. After taking into account the uncertainties of the input parameters, we obtain the  hadronic coupling constants, which are shown explicitly in Table 3. Now it is straightforward to calculate the partial decay widths of the $Y(4660)\to J/\psi f_0(980)$, $ \eta_c \phi(1020)$,    $ \chi_{c0}\phi(1020)$, $ D_s \bar{D}_s$, $ D_s^* \bar{D}^*_s$, $ D_s \bar{D}^*_s$,  $ D_s^* \bar{D}_s$ with formula,
\begin{eqnarray}
\Gamma\left(Y(4660)\to BC\right)&=& \frac{p(m_Y,m_B,m_C)}{24\pi m_Y^2} |T^2|\, ,
\end{eqnarray}
where $p(a,b,c)=\frac{\sqrt{[a^2-(b+c)^2][a^2-(b-c)^2]}}{2a}$, the $T$ are the scattering amplitudes defined in Eq.(39), the numerical values of the
partial decay widths are shown in Table 3.

\begin{figure}
 \centering
  \includegraphics[totalheight=5cm,width=7cm]{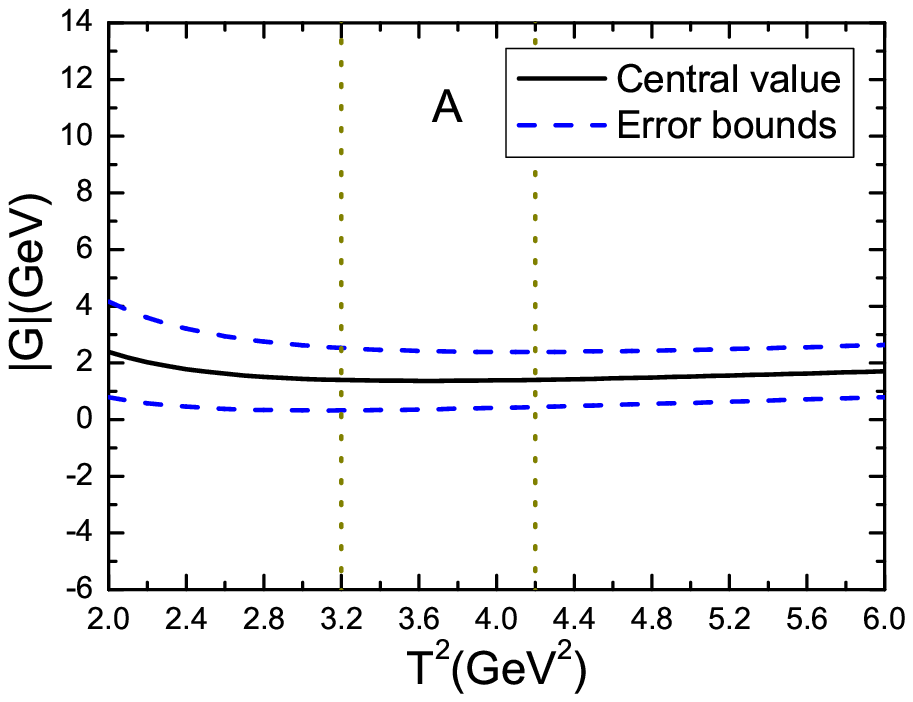}
  \includegraphics[totalheight=5cm,width=7cm]{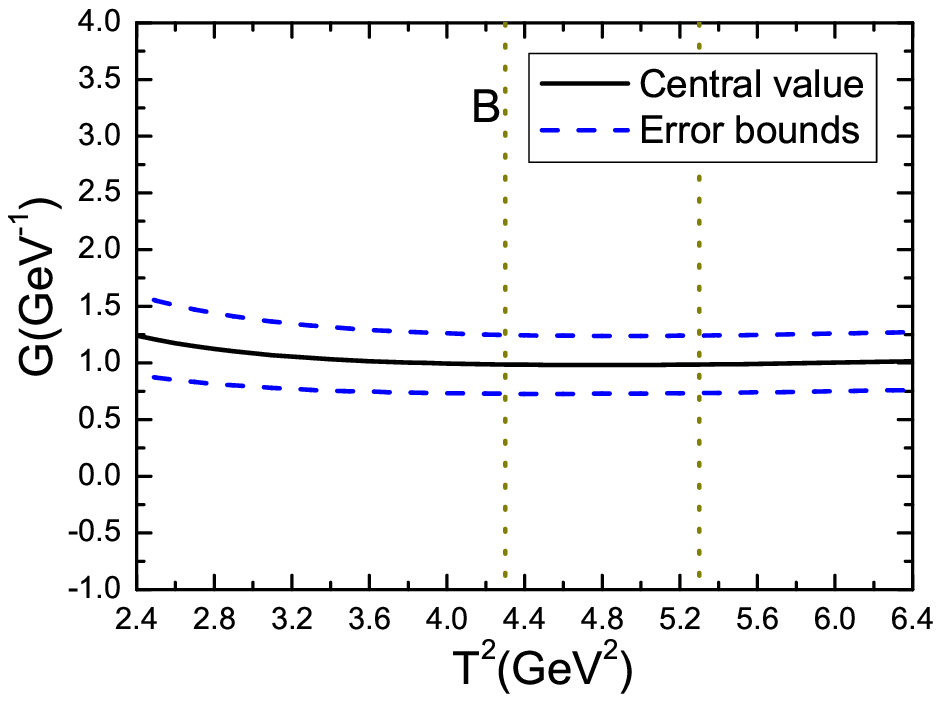}
   \includegraphics[totalheight=5cm,width=7cm]{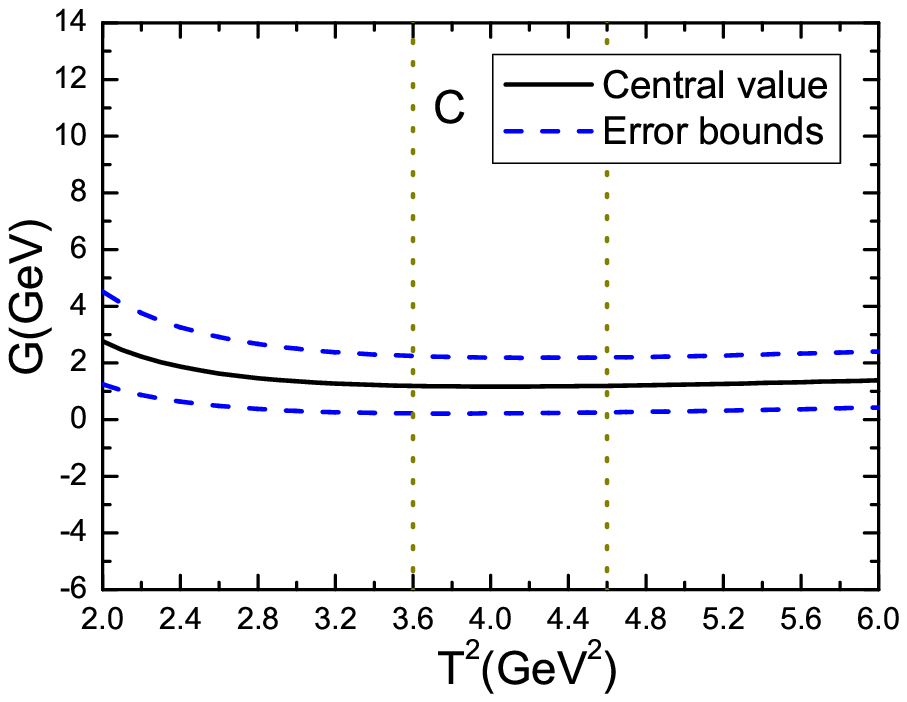}
  \includegraphics[totalheight=5cm,width=7cm]{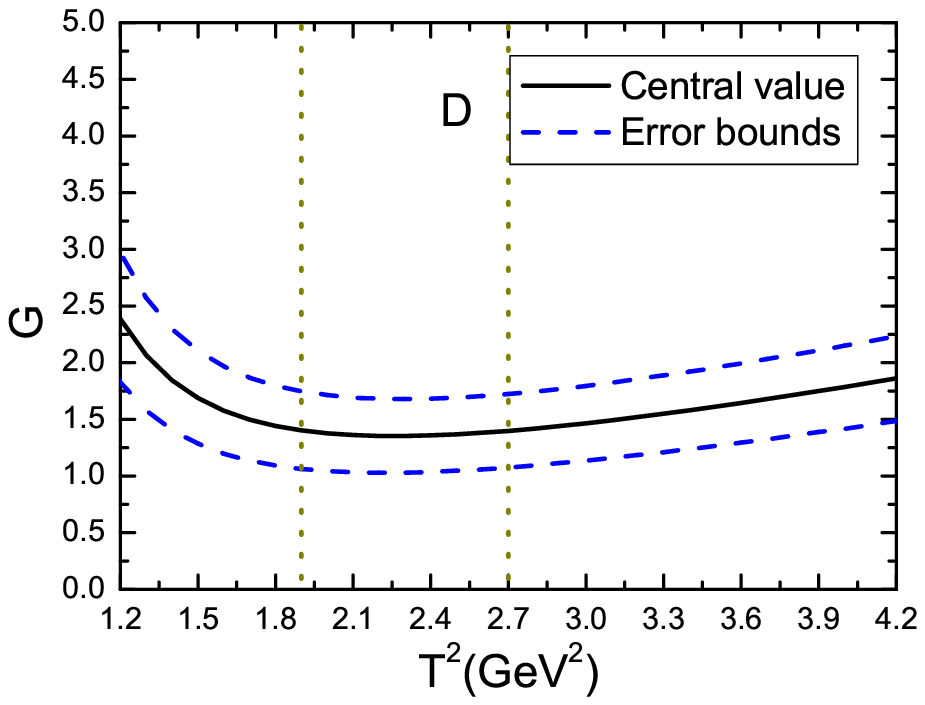}
   \includegraphics[totalheight=5cm,width=7cm]{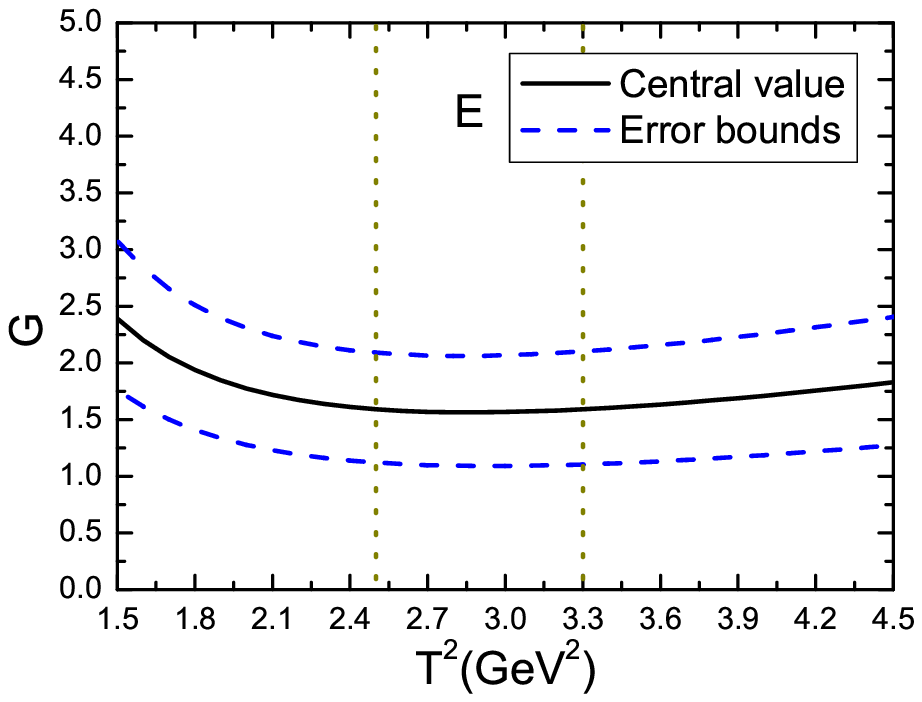}
     \includegraphics[totalheight=5cm,width=7cm]{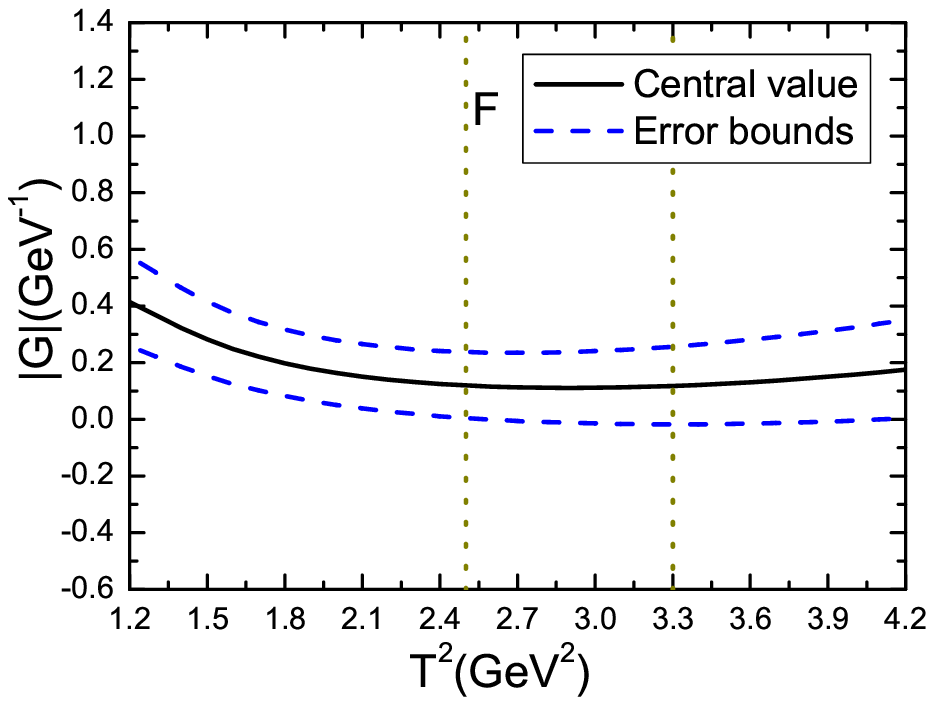}
     \caption{The hadronic coupling constants  with variations of the  Borel  parameters  $T^2$, where the $A$, $B$, $C$, $D$, $E$ and $F$ denote the $G_{YJ/\psi f_0}$,
     $ G_{Y\eta_c \phi}$, $G_{Y\chi_{c0} \phi}$, $G_{YD_s\bar{D}_s}$, $G_{YD_s^*\bar{D}_s^*}$ and $G_{YD_s\bar{D}_s^*}$, respectively, the regions between the two
     perpendicular lines are the Borel windows.}
\end{figure}

\begin{figure}
 \centering
  \includegraphics[totalheight=5cm,width=7cm]{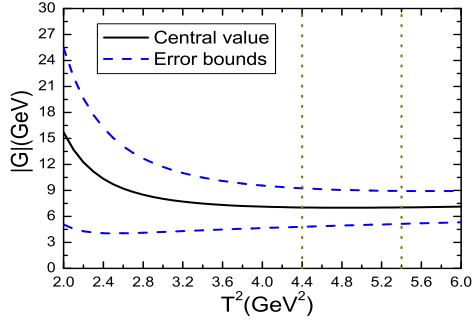}
     \caption{The hadronic coupling constant $G_{Y\psi^\prime f_0}$ with variation of the  Borel  parameter $T^2$, the region between the two
     perpendicular lines is the Borel window.}
\end{figure}

The decay $Y(4660)\to \psi^{\prime}f_0(980)$ is   kinematically forbidden, but the decay  $Y(4660)\to \psi^{\prime}\pi^+\pi^-$ can take place through a  virtual intermediate $f_0(980)^*$, the partial decay width can be written as,
\begin{eqnarray}
\Gamma(Y\to \psi^\prime\pi^+\pi^-)&=&\int_{4m_\pi^2}^{(m_Y-m_{\psi^\prime})^2}ds\,|T|^2\frac{p(m_Y,m_{\psi^\prime},\sqrt{s})\,p(\sqrt{s},m_\pi,m_\pi)}{192\pi^3 m_Y^2 \sqrt{s} }\ , \nonumber\\
&=&5.5^{+4.1}_{-2.9}\,\rm{MeV}\, ,
\end{eqnarray}
where
\begin{eqnarray}
|T|^2&=&\frac{(M_Y^2-s)^2+2(5M_Y^2-s)M^2_{\psi^\prime}+M^4_{\psi^\prime}}{4M_Y^2 M_{\psi^\prime}^2}G_{Y\psi^{\prime}f_0}^2\frac{1}{(s-m_{f_0}^2)^2+s\Gamma_{f_0}^2(s)}G_{f_0\pi\pi}^2\, , \nonumber\\
\Gamma_{f_0}(s)&=&\Gamma_{f_0}(m_{f_0}^2) \frac{m_{f_0}^2}{s}\sqrt{\frac{s-4m_{\pi}^2}{m_{f_0}^2-4m_{\pi}^2}}\, , \nonumber\\
\Gamma_{f_0}(m_{f_0}^2)&=& \frac{G_{f_0\pi\pi}^2}{16\pi m_{f_0}^2}\sqrt{ m_{f_0}^2-4m_{\pi}^2 }\, ,
\end{eqnarray}
 $\Gamma_{f_0}(m_{f_0}^2)=50\,\rm{MeV}$ \cite{PDG}, the hadronic coupling constant $G_{f_0\pi\pi}$ is  defined by $\langle \pi^+(p)\pi^-(q)|f_0(p^\prime)\rangle=iG_{f_0\pi\pi}$.

Now it is easy to obtain the total decay width,
\begin{eqnarray}
\Gamma(Y(4660) )&=& 74.2^{+29.2}_{-19.2}\,{\rm{MeV}}\, .
\end{eqnarray}
	The predicted width $\Gamma(Y(4660) )= 74.2^{+29.2}_{-19.2}\,{\rm{MeV}}$ is in excellent agreement with the experimental data $68\pm 11\pm 1 {\mbox{ MeV}}$ from the Belle collaboration \cite{Belle-Y4660-2014}, which also supports assigning the $Y(4660)$ to be the  $[sc]_P[\bar{s}\bar{c}]_A-[sc]_A[\bar{s}\bar{c}]_P$  type tetraquark state with $J^{PC}=1^{--}$.

From Table 3, we can see that the  hadronic coupling constants $ |G_{Y\psi^\prime f_0}|=7.00^{+2.24}_{-2.20}\,{\rm{GeV}}
\gg |G_{Y J/\psi f_0}|=1.37^{+1.16}_{-1.04}\,\rm{GeV}$, which indicates that the coupling  $Y(4660)\psi^{\prime}f_0(980)$  is very strong, and  consistent with the observation of the $Y(4660)$ in the $\psi^\prime\pi^+\pi^-$ mass spectrum, and favors the $\psi^{\prime}f_0(980)$ molecule assignment \cite{FKGuo-4660,Wang-CTP-4660,Nielsen-4660-mole}, as the strong coupling maybe lead to  some $\psi^{\prime}f_0(980)$ component.
Now we perform Fierz re-arrangement  to the vector current $J_{\mu}(x)$ both in the color and Dirac-spinor  spaces,  and obtain the  result,
\begin{eqnarray}
J_{\mu} &=&\frac{1}{2\sqrt{2}}\Big\{\,\bar{c} \gamma^\mu c\,\bar{s} s-\bar{c} c\,\bar{s}\gamma^\mu s+i\bar{c}\gamma^\mu\gamma_5 s\,\bar{s}i\gamma_5 c-i\bar{c} i\gamma_5 s\,\bar{s}\gamma^\mu \gamma_5c  \nonumber\\
&&  - i\bar{c}\gamma_\nu\gamma_5c\, \bar{s}\sigma^{\mu\nu}\gamma_5s+i\bar{c}\sigma^{\mu\nu}\gamma_5c\, \bar{s}\gamma_\nu\gamma_5s
- i\bar{s}\gamma_\nu c\, \bar{c}\sigma^{\mu\nu}s+i \bar{s}\sigma^{\mu\nu}c \,\bar{c}\gamma_\nu s  \,\Big\} \, .
\end{eqnarray}
The $J_{\mu}(x)$ can be taken  as a special superposition of color singlet-singlet type currents,  which couple potentially  to the meson-meson pairs or molecular states. The first term $\bar{c} \gamma^\mu c\,\bar{s} s$ is the molecule current chosen in Refs.\cite{Wang-CTP-4660,Nielsen-4660-mole}, which couples potentially to the $\psi^{\prime}f_0(980)$ molecular state. There does not  exist a term $\bar{c}\sigma_{\mu\nu}c\, \bar{s}\gamma^{\nu}s$, which couples potentially  to the $J/\psi \phi(1020)$ or $\psi^\prime \phi(1020)$ molecular state or scattering state. In calculations, we observe that
the   QCD side of the component  $\Pi(p^{\prime2},p^2,q^2)$ in the correlation function  $\Pi_{\alpha\beta\nu}(p,q)$ in Eq.(37)  is zero at the leading order approximation, the hadronic coupling constant $G_{YJ/\psi\phi}\approx 0$. The decay $Y(4660)\to J/\psi \phi(1020)$ is greatly suppressed and can take place only through rescattering mechanism.
It is important to search for the process $Y(4660)\to J/\psi \phi(1020)$ to diagnose the structure of the $Y(4660)$.

In Ref.\cite{Azizi-4660}, Sundu, Agaev and Azizi choose  the $[sc]_S[\bar{s}\bar{c}]_V+[sc]_V[\bar{s}\bar{c}]_S$ type current to study the mass and width of the $Y(4660)$, and obtain the values $ m_Y=4677^{+71}_{-63}\ \mathrm{MeV}$ and $\Gamma_{Y}= (64.8 \pm 10.8)\ \mathrm{MeV}$ by saturating the width with the decays $Y(4660) \to J/\psi f_0(500)$, $J/\psi f_0(980)$, $\psi^\prime f_0(500)$, $\psi^\prime f_0(980)$. If the experimental value $m_{Y}=4652\pm10\pm 8\,\rm{MeV}$ is taken, the decay $Y(4660)\to \psi^\prime f_0(980)$ is kinematically forbidden, and can only take place through the upper tail of the mass distribution, the prediction $\Gamma(Y(4660)\to \psi^{\prime}f_0(980))=30.2 \pm 8.5\,\rm{MeV}$ is too large. Furthermore, other decay channels should be taken into account.

\section{Conclusion}
In this article, we illustrate how to calculate the hadronic coupling constants in the strong decays of the tetraquark states based on solid quark-hadron quality, then
study the hadronic coupling constants $G_{Y J/\psi f_0}$, $G_{Y\eta_c \phi}$,    $G_{Y\chi_{c0}\phi}$, $G_{YD_s \bar{D}_s}$, $G_{Y D_s^* \bar{D}^*_s}$,
$ G_{YD_s \bar{D}^*_s}$,   $G_{Y\psi^\prime f_0}$, $G_{YJ/\psi\phi}$ in the decays $Y(4660)\to J/\psi f_0(980)$, $ \eta_c \phi(1020)$,    $ \chi_{c0}\phi(1020)$, $ D_s \bar{D}_s$, $ D_s^* \bar{D}^*_s$, $ D_s \bar{D}^*_s$,   $  \psi^\prime \pi^+\pi^-$, $J/\psi\phi(1020)$ with the QCD sum rules in a systematic way. The predicted width $\Gamma(Y(4660) )= 74.2^{+29.2}_{-19.2}\,{\rm{MeV}}$ is in excellent agreement with the experimental data $68\pm 11\pm 1 {\mbox{ MeV}}$ from the Belle collaboration, which supports assigning the $Y(4660)$ to be the  $[sc]_P[\bar{s}\bar{c}]_A-[sc]_A[\bar{s}\bar{c}]_P$  type tetraquark state with $J^{PC}=1^{--}$.
In calculations, we observe that the  hadronic coupling constants $ |G_{Y\psi^\prime f_0}|\gg |G_{Y J/\psi f_0}|$, which indicates that the coupling  $Y(4660)\psi^{\prime}f_0(980)$  is very strong, and  consistent with the observation of the $Y(4660)$ in the $\psi^\prime\pi^+\pi^-$ mass spectrum, and favors the $\psi^{\prime}f_0(980)$ molecule assignment, as there may be appear some $\psi^{\prime}f_0(980)$ component due to the strong coupling. The decay $Y(4660)\to J/\psi \phi(1020)$ is greatly suppressed and can take place only through rescattering mechanism. It is important to search for the process $Y(4660)\to J/\psi \phi(1020)$ to diagnose the nature  of the $Y(4660)$.

\section*{Acknowledgements}
This  work is supported by National Natural Science Foundation, Grant Number  11775079.

\end{document}